\documentclass[10pt]{article}

\usepackage[margin=1.2in]{geometry}  
\usepackage{graphicx}              
\usepackage{amsmath}               
\usepackage{amsfonts}              
\usepackage{amsthm}                

\usepackage{subfigure}             

\begin{document}


\title{\Large \bf Space-time goal-oriented reduced basis approximation for linear wave equation}

\author{K. C. Hoang, P. Kerfriden and S. P. A. Bordas\thanks{Email: \{hoangkhacchi, pierre.kerfriden, stephane.bordas\}@gmail.com} \vspace{9pt}\\
Cardiff University, School of Engineering, \\
The Parade, CF24 3AA Cardiff, United Kingdom }

\date{}

\maketitle

\begin{abstract}

\noindent In this paper, we study numerically the linear damped second-order hyperbolic partial differential equation (PDE) with affine parameter dependence using a goal-oriented approach by finite element (FE) and reduced basis (RB) methods. The main contribution of this paper is the ``goal-oriented'' proper orthogonal decomposition (POD)--Greedy sampling procedure within the RB approximation context. First, we introduce the RB recipe: Galerkin projection onto a space $Y_N$ spanned by solutions of the governing PDE at $N$ selected points in parameter space. This set of $N$ parameter points is constructed by the standard POD--Greedy sampling procedure already developed. Second, based on the affine parameter dependence, we make use of the offline-online computational procedures: in the offline stage, we generate the RB space; in the online stage, given a new parameter value, we calculate rapidly and accurately the space-time RB output of interest and its associated asymptotic error. The proposed goal-oriented POD--Greedy sampling procedure can now be implemented and will look for the parameter points such that it minimizes this (asymptotic) output error rather than the solution error (or, error indicator which is the dual norm of residual) as in the standard POD--Greedy procedure. Numerical results show that the new goal-oriented POD--Greedy sampling procedure improves significantly the accuracy of the space-time output computation in comparison with the standard POD--Greedy one. The method is thus ideally suited for repeated, rapid and reliable evaluation of input-output relationships within the space-time setting.

\bigskip


\end{abstract}

\section{Introduction}\label{sec_Intro}

The design, optimization and control procedures of engineering problems often require several forms of performance measures or outputs -- such as displacements, heat fluxes or flowrates \cite{Grepl2005}. Generally, these outputs are functions of field variables such as displacements, temperature or velocities which are usually governed by a PDE. The parameter or input will frequently define a particular configuration of the model problem. Therefore, the relevant system behavior will be described by an implicit input-output relationship; where its computation requires the solution of the underlying parameter-PDE (or $\mu$PDE). We pursue the RB method \cite{Rozza2008, Hoang2013} which permits the \textit{efficient} and \textit{reliable} evaluation of this PDE-induced input-output relationship in \textit{many query} and \textit{real-time} contexts.

The RB method was first introduced in the late 1970s for nonlinear analysis of structures and has been further investigated and developed more broadly. Recently, the RB method was well developed for various kinds and classes of parametrized PDEs such as: the eigenvalue problems, the coercive/non-coercive affine/non-affine linear/nonlinear elliptic PDEs, the coercive/non-coercive affine/non-affine linear/nonlinear parabolic PDEs, the coercive affine linear hyperbolic PDEs, and several highly nonlinear problems such as Burger's equation and Boussinesq equation. All of these works, which were proposed and performed by Patera and co-workers, can be found under an abstract list form in the website \cite{PateraWebsite}. For the linear wave equation, the RB method and associated \textit{a posteriori} error estimation was developed successfully with some levels \cite{Tan2007, Huynh2011b, HoangPhDThesis2012}; however, none of these works have focused on goal-oriented RB approximation and its associated error estimation.

Adaptive finite element (FE) methods and goal-oriented error estimates for the wave equation have been investigated widely in many applications \cite{Gratsch2005}. Among those methods, the most well-known one is the dual-weighted residual (DWR) method which was proposed by Rannacher and co-workers \cite{Bangerth1999, Bangerth2001, Becker2001, Bangerth2010}. In those works, the authors have used the DWR method to quantify the \textit{a posteriori} error of the interest output in order to finer locally the finite element mesh in an adaptive manner. The final goal is to minimize computational efforts and maximize the accuracy of the interest output in an adaptive and controllable manner. In particular, the DWR method makes use of an auxiliary dual (or sensitivity) equation to derive an \textit{a posteriori} error expression for the interest output from the primal residual and the dual solution of that dual equation in space-time setting \cite{Gratsch2005}. The name ``dual-weighted residual'' is thus derived from this fact.

It is thus very natural to combine goal-oriented estimation with RB approximations \cite{Grepl2005, Meyer2003} to improve the accuracy of the interest output within RB context. Bringing the ``DWR idea'' to the RB context, we can also obtain an error expression which relates the output error with the primal residual of the dual solution in space-time setting. Namely, the error between the (space-time) FE and RB outputs will be exactly equal to the primal residual of the FE dual solution \cite{Bangerth2010}. However, to make this error expression truly \textit{a posteriori}, ones prefer to approximate the FE dual solution by some RB ones rather than using it directly within the expression \cite{Grepl2005, Meyer2003}. Hence, the error expression now becomes an \textit{a posteriori} error approximation since the equal sign (in the expression) now becomes the approximation one. This approach is generally good; but it suffers from two main drawbacks: first, it is only an error approximation which is not a true upper error bound and second, underlying concepts and computational effort are high/complicated since we need to solve additionally a dual problem and all related issues. We also implemented this approach recently in the work \cite{Hoang2013acme}.

Here, we also give a brief review on goal-oriented estimations within reduced order modeling context using the snapshots-POD method \cite{Sirovich1987, Kerfriden2010, Kerfriden2011a, Kerfriden2011b, Kerfriden2012}. Goal-oriented estimations using the snapshots-POD method were implemented in \cite{Willcox2005, Bui2007}. In this approach, the authors solved an PDE-constrained optimization problem to find the optimal set of basis functions. In particular, the optimal basis functions are found such that they minimize the true output errors (with appropriate regularization techniques) and subject to equilibrium PDE-constraints \cite{Nocedal1999, Heinkenschloss2008}. This approach is optimal, however, it is very expensive since ones have to compute all the FEM solutions/outputs in every iteration within optimization solvers; and hence, it would limit the number of input parameters in comparison with the RB approach.

In this paper, we devise a simpler approach that addresses the two aforementioned drawbacks. First, we use asymptotic output errors which are also not truly upper error bounds but asymptotically converged to the true error; their performances are well comparable with the \textit{a posteriori} DWR error expression. Second, the approach completely do not use any information of the dual problem and hence simpler and easier to implement. Based on the standard POD--Greedy sampling procedure and the asymptotic output errors, we propose the new ``goal-oriented'' POD--Greedy sampling procedure, which will pick up the parameter points such that the error (or error approximation) of the output functional is minimized. This idea is novel and further developed the idea of the standard POD--Greedy sampling procedure currently used \cite{Haasdonk2008, Knezevic2011, Hoang2013}, where the algorithm will pick up optimally all parameter points such that the error (or error indicator) of the field variable is minimized. By this way, we expect to improve significantly the accuracy of the RB output functional computations; but consequently, we might lose the rapid convergent rate of the field variable as in the standard POD--Greedy algorithm. In fact, as we can see later in the numerical results section, the convergent rate of the field variable by the two algorithms are quite similar; while the convergent rate of the output by the goal-oriented POD--Greedy algorithm is faster than that of the standard POD--Greedy one\footnotemark \footnotetext{In subsequent sections, for simplicity we shall call the ``standard algorithm'' to mention the standard POD--Greedy algorithm, and the ``goal-oriented algorithm'' to mention the goal-oriented POD--Greedy algorithm, respectively.}.

The paper is organized as follows. In Section \ref{sec_ProbStat}, we introduce necessary definitions, concepts and notations and then state the problem using a semidiscrete approach: fully discretizing in space using Galerkin FEM and marching in time using Newmark's trapezoidal rule. In Section \ref{sec_RB}, we describe various topics related to the RB methodology: approximation, the standard versus goal-oriented algorithms, error estimations and offline-online computational procedure. We then present some numerical results of the three-dimensional dental implant problem \cite{Hoang2013} to verify the performance of the proposed algorithm in Section \ref{sec_NumEx}. Finally, we provide some concluding remarks in Section \ref{sec_Conclusion}.

\section{Problem statement}\label{sec_ProbStat}

\subsection{Abstract formulation}\label{subsec_ProbStat_AbstForm}

We consider a spatial domain $\Omega \in \mathbb{R}^d$ with Lipschitz continuous boundary $\partial \Omega$. We denote the Dirichlet portion of the boundary by $\Gamma_i^D, 1 \le i \le d$. We then introduce the Hilbert spaces

\begin{subequations}\label{eq:ProbStat_one}
\begin{equation}
Y^e = \{ v \equiv (v_1,\ldots,v_d) \in (H^1(\Omega))^d \; | \; v_i=0 \quad {\rm on} \quad \Gamma^D_i, i=1,\ldots,d \}, \label{eq:ProbStat_one-a}
\end{equation}
\begin{equation}
X^e = (L^2(\Omega))^d. \label{eq:ProbStat_one-b}
\end{equation}
\end{subequations}

Here, $H^1(\Omega) = \{ v \in L^2(\Omega) \; | \; \nabla v \in (L^2(\Omega))^d \}$ where $L^2(\Omega)$ is the space of square-integrable functions over $\Omega$. We equip our spaces with inner products and associated norms $(\cdot,\cdot)_{Y^e}$ $((\cdot,\cdot)_{X^e})$ and $\|\cdot\|_{Y^e}=\sqrt{(\cdot,\cdot)_{Y^e}}$ $(\|\cdot\|_{X^e}=\sqrt{(\cdot,\cdot)_{X^e}})$, respectively; a typical choice is

\begin{subequations}\label{eq:ProbStat_two}
\begin{equation}
(w,v)_{Y^e} = \int_{\Omega} \frac{\partial w_i}{\partial x_j} \frac{\partial v_i}{\partial x_j} + w_i v_i, \label{eq:ProbStat_two-a}
\end{equation}
\begin{equation}
(w,v)_{X^e} = \int_{\Omega} w_i v_i, \label{eq:ProbStat_two-b}
\end{equation}
\end{subequations}

\noindent where the summation over repeated indices is assumed.

We next define our parameters set $\mathcal{D} \in \mathbb{R}^P$, a typical point in which shall be denoted $\mu \equiv (\mu_1,\ldots,\mu_P)$. We then define the parametrized bilinear forms $a$ in $Y^e$, $a:Y^e \times Y^e \times \mathcal{D} \rightarrow \mathbb{R}$; $m,c,f,\ell$ are continuous bilinear and linear forms in $X^e$, $m: X^e \times X^e \times \mathcal{D} \rightarrow \mathbb{R}$, $c:X^e \times X^e \times \mathcal{D} \rightarrow \mathbb{R}$, $f: X^e \times \mathcal{D} \rightarrow \mathbb{R}$ and $\ell: X^e \rightarrow \mathbb{R}$.

For simplicity, we directly consider a time-discrete framework associated to the time interval $I = [0,T]$. We divide $I$ into $K$ subintervals of equal length $\Delta t = \frac{T}{K}$ and define $t^k = k \Delta t$, $0 \le k \le K$. We shall consider the Newmark's trapezoidal scheme with coefficients $\left( \gamma_s=\frac{1}{2}, \beta_s=\frac{1}{4} \right)$ for the time integration. Our results must be stable as $\Delta t \rightarrow 0, K \rightarrow \infty$.

The ``exact'' semi-discrete problem is stated follows: given a parameter $\mu \in \mathcal{D} \subset \mathbb{R}^{P}$, we evaluate the (space-time) output of interest

\begin{equation}\label{eq:ProbStat_three}
s(\mu) = \sum_{k=0}^{K-1} \int_{t^k}^{t^{k+1}} \int_{\Gamma_{\rm o}} u^e(x,t;\mu) \, \Sigma(x,t) dx dt = \sum_{k=0}^{K-1} \int_{t^k}^{t^{k+1}} \ell(u^e(x,t;\mu)) dt,
\end{equation}

\noindent where the field variable, $u^e(\mu,t^k) \in Y^e$, $1 \le k \le K$, satisfies the weak form of the $\mu$-parametrized hyperbolic PDE \cite{Daniel1997}

\begin{multline}\label{eq:ProbStat_four}
\frac{1}{\Delta t^2} m(u^e(\mu,t^{k+1}),v;\mu) + \frac{1}{2 \Delta t} c(u^e(\mu,t^{k+1}),v;\mu) + \frac{1}{4} a(u^e(\mu,t^{k+1}),v;\mu) \\
= - \frac{1}{\Delta t^2} m(u^e(\mu,t^{k-1}),v;\mu) + \frac{1}{2 \Delta t} c(u^e(\mu,t^{k-1}),v;\mu) - \frac{1}{4} a(u^e(\mu,t^{k-1}),v;\mu) \\
+ \frac{2}{\Delta t^2} m(u^e(\mu,t^{k}),v;\mu) - \frac{1}{2} a(u^e(\mu,t^{k}),v;\mu) + g^{eq}(t^k) f(v;\mu), \quad \forall v \in Y^e, 1 \leq k \leq K-1,
\end{multline}

\noindent with

\begin{equation}\label{eq:ProbStat_five}
g^{eq}(t^k) = \frac{1}{4} g(t^{k-1}) + \frac{1}{2} g(t^k) + \frac{1}{4} g(t^{k+1}), \quad 1 \le k \le K-1,
\end{equation}

\noindent and initial conditions: $u^e(\mu,t^0)=0$, $\frac{\partial u^e(\mu,t^0)}{\partial t}=0$.

Here, $\Gamma_{\rm o}$ is some (output) spatial regions of interest and $\Sigma(x,t)$ is an extractor which depends on the view position of an ``observer'' in the space-time domain; and $\ell(u^e(x,t;\mu)) = \int_{\Gamma_{\rm o}} u^e(x,t;\mu) \, \Sigma(x,t) dx$.

We next introduce a reference finite element approximation space $Y \subset Y^e (\subset X^e)$ of dimension $\mathcal{N}$; we further define $X \equiv X^e$. Note that $Y$ and $X$ shall inherit the inner product and norm from $Y^e$ and $X^e$, respectively. Our ``true'' finite element approximation $u(\mu,t^k) \in Y$ to the ``exact'' problem is stated as

\begin{multline}\label{eq:ProbStat_six}
\frac{1}{\Delta t^2} m(u(\mu,t^{k+1}),v;\mu) + \frac{1}{2 \Delta t} c(u(\mu,t^{k+1}),v;\mu) + \frac{1}{4} a(u(\mu,t^{k+1}),v;\mu) \\
= - \frac{1}{\Delta t^2} m(u(\mu,t^{k-1}),v;\mu) + \frac{1}{2 \Delta t} c(u(\mu,t^{k-1}),v;\mu) - \frac{1}{4} a(u(\mu,t^{k-1}),v;\mu) \\
+ \frac{2}{\Delta t^2} m(u(\mu,t^{k}),v;\mu) - \frac{1}{2} a(u(\mu,t^{k}),v;\mu) + g^{eq}(t^k) f(v;\mu), \quad \forall v \in Y, 1 \leq k \leq K-1,
\end{multline}

\noindent with initial conditions\footnotemark \footnotetext{In order to start the procedure \eqref{eq:ProbStat_six}, $u(\mu,t^1)$ is computed as on page 491 of \cite{Hughes1987}.}: $u(\mu,t^0)=0$, $\frac{\partial u(\mu,t^0)}{\partial t}=0$; we then evaluate the interest output from

\begin{equation}\label{eq:ProbStat_seven}
s(\mu) = \sum_{k=0}^{K-1} \int_{t^k}^{t^{k+1}} \ell(u(x,t;\mu)) dt.
\end{equation}

The RB approximation shall be built upon our reference finite element approximation, and the RB error will thus be evaluated with respect to $u(\mu,t^k) \in Y$. Clearly, our methods must remain computationally efficient and stable as $\mathcal{N} \rightarrow \infty$.

We shall make the following assumptions. First, we assume that the bilinear forms $a(\cdot,\cdot;\mu)$ and $m(\cdot,\cdot;\mu)$ are continuous,

\begin{subequations}\label{eq:ProbStat_eight}
\begin{equation}
a(w,v;\mu) \leq \gamma\|w\|_Y \|v\|_Y \leq \gamma_0\|w\|_Y \|v\|_Y, \quad \forall w,v \in Y, \forall \mu \in \mathcal{D},    \label{eq:ProbStat_eight-a}
\end{equation}
\begin{equation}
m(w,v;\mu) \leq \rho\|w\|_X \|v\|_X \leq \rho_0\|w\|_X \|v\|_X, \quad \forall w,v \in Y, \forall \mu \in \mathcal{D},    \label{eq:ProbStat_eight-b}
\end{equation}
\end{subequations}

\noindent coercive,

\begin{subequations}\label{eq:ProbStat_nine}
\begin{equation}
0 \leq \alpha_0 \leq \alpha(\mu) \equiv \inf_{v \in Y} \frac{a(v,v;\mu)}{\|v\|^2_Y}, \quad \forall \mu \in \mathcal{D},    \label{eq:ProbStat_nine-a}
\end{equation}
\begin{equation}
0 \leq \sigma_0 \leq \sigma(\mu) \equiv \inf_{v \in Y} \frac{m(v,v;\mu)}{\|v\|^2_X}, \quad \forall \mu \in \mathcal{D};    \label{eq:ProbStat_nine-b}
\end{equation}
\end{subequations}

\noindent and symmetric $a(v,w;\mu)=a(w,v;\mu), \forall w,v \in Y, \forall \mu \in \mathcal{D}$, and $m(v,w;\mu)=m(w,v;\mu), \forall w,v \in X, \forall \mu \in \mathcal{D}$. (We (plausibly) suppose that $\gamma_0, \rho_0, \alpha_0$ and $\sigma_0$ may be chosen independent of $\mathcal{N}$ \cite{Grepl2005}). We also require that the linear forms $f(\cdot;\mu): Y \rightarrow \mathbb{R}$ and $\ell(\cdot): Y \rightarrow \mathbb{R}$ be bounded with respect to $\|\cdot\|_Y$ and $\|\cdot\|_X$, respectively.

Second, we shall assume that $a$, $m$, $c$ and $f$ depend affinely on the parameter $\mu$ and thus can be expressed as

\begin{subequations}\label{eq:ProbStat_ten}
\begin{equation}
m(w,v;\mu) = \sum_{q=1}^{Q_m} \Theta^q_m(\mu)m^q(w,v), \quad \forall w,v \in Y, \mu \in \mathcal{D}, \label{eq:ProbStat_ten-a}
\end{equation}
\begin{equation}
c(w,v;\mu) = \sum_{q=1}^{Q_c} \Theta^q_c(\mu)c^q(w,v), \quad \forall w,v \in Y, \mu \in \mathcal{D}, \label{eq:ProbStat_ten-b}
\end{equation}
\begin{equation}
a(w,v;\mu) = \sum_{q=1}^{Q_a} \Theta^q_a(\mu)a^q(w,v), \quad \forall w,v \in Y, \mu \in \mathcal{D}, \label{eq:ProbStat_ten-c}
\end{equation}
\begin{equation}
f(v;\mu) = \sum_{q=1}^{Q_f} \Theta^q_f(\mu)f^q(v), \quad \forall v \in Y, \mu \in \mathcal{D}, \label{eq:ProbStat_ten-d}
\end{equation}
\end{subequations}

\noindent for some (preferably) small integers $Q_{m,c,a,f}$. Here, the smooth functions $\Theta^q_{m,c,a,f}(\mu): \mathcal{D} \rightarrow \mathbb{R}$ depend on $\mu$, but the bilinear and linear forms $m^q$, $c^q$, $a^q$ and $f^q$ do \textit{not} depend on $\mu$.

Finally, we also require that all linear and bilinear forms be independent of time -- the system is thus linear time-invariant (LTI) \cite{Grepl2005}. We shall point out that one application which satisfies this assumption is the dental implant problem \cite{Hoang2013, Zaw2009}.

%
%

\subsection{Impulse response}\label{subsec_ProbStat_ImpRes}

In many dynamical systems, generally, the applied force to excite the system (e.g., $g(t^k)$ in \eqref{eq:ProbStat_five}) is not known in advance and thus we cannot solve \eqref{eq:ProbStat_six} for $u(\mu,t^k)$. In such situations, fortunately, we may appeal to the LTI hypothesis to justify an impulse approach as described now \cite{Grepl2005}. We note from the Duhamel's principle that the solution of any LTI system can be written as the convolution of the impulse response with the control input: for any control input $g^{\rm any}(t^k)$, we can obtain its corresponding solution $u^{\rm any}(\mu,t^k)$, $1 \leq k \le K$ from

\begin{equation}\label{eq:ProbStat_eleven}
u^{\rm any}(\mu,t^k)=\sum_{j=1}^{k} u^{\rm unit}(\mu,t^{k-j+1}) g^{\rm any}(t^j), \quad 1 \le k \le K,
\end{equation}

\noindent where $u^{\rm unit}(\mu,t^k)$ is the solution of \eqref{eq:ProbStat_six} for a unit impulse control input $g^{\rm unit}(t^k)=\delta_{1k}$, $1 \le k \le K$ ($\delta$ is the Kronecker delta symbol). Therefore, it is sufficient to build the RB basis functions for the problem based on this impulse response \cite{Grepl2005}.

\section{Reduced basis approximation}\label{sec_RB}

\subsection{Approximation}\label{subsec_RB_Apprx}

We introduce the nested samples $S_* = \{\mu_1 \in \mathcal{D}, \mu_2 \in \mathcal{D}, \ldots, \mu_N \in \mathcal{D}\}, 1 \le N \le N_{\max} $, and associated nested Lagrangian RB spaces $Y_N = {\rm span} \{\zeta_n, 1 \le n \le N\}, 1 \le N \le N_{\max}$, where $\zeta_n \in Y_N, 1 \le n \le N_{\max}$ are mutually $(\cdot,\cdot)_Y$ -- orthogonal RB basis functions. The sets $S_*$ and $Y_N$ shall be constructed appropriately by the standard and goal-oriented POD--Greedy algorithms described in Section \ref{subsec_RB_GOPODGreedy} afterward.

Our reduced basis approximation $u_N(\mu,t^k)$ to $u(\mu,t^k)$ is then obtained by a standard Galerkin projection: given $\mu \in \mathcal{D}$, we now look for $u_N(\mu,t^k) \in Y_N$ satisfies

\begin{multline}\label{eq:RB_one}
\frac{1}{\Delta t^2} m(u_N(\mu,t^{k+1}),v;\mu) + \frac{1}{2 \Delta t} c(u_N(\mu,t^{k+1}),v;\mu) + \frac{1}{4} a(u_N(\mu,t^{k+1}),v;\mu) \\
= - \frac{1}{\Delta t^2} m(u_N(\mu,t^{k-1}),v;\mu) + \frac{1}{2 \Delta t} c(u_N(\mu,t^{k-1}),v;\mu) - \frac{1}{4} a(u_N(\mu,t^{k-1}),v;\mu) \\
+ \frac{2}{\Delta t^2} m(u_N(\mu,t^{k}),v;\mu) - \frac{1}{2} a(u_N(\mu,t^{k}),v;\mu) + g^{eq}(t^k) f(v;\mu), \quad \forall v \in Y_N, 1 \leq k \leq K-1,
\end{multline}

\noindent where the zero initial conditions are defined and treated as mentioned in Section \ref{subsec_ProbStat_AbstForm}; we then evaluate the output estimate, $s_N(\mu)$, from

\begin{equation}\label{eq:RB_two}
s_N(\mu) = \sum_{k=0}^{K-1} \int_{t^k}^{t^{k+1}} \ell(u_N(x,t;\mu)) dt.
\end{equation}

\subsection{Goal-oriented POD--Greedy sampling procedure}\label{subsec_RB_GOPODGreedy}

\subsubsection{The proper orthogonal decomposition}

We aim to generate an optimal (in the mean square error sense) basis set $\{ \zeta_m \}_{m=1}^{M}$ from any given set of $M_{\max} (\ge M)$ snapshots $\{\xi_k\}_{k=1}^{M_{\max}}$. To do this, let $V_M = {\rm span} \{v_1, \ldots, v_M\} \subset {\rm span} \{\xi_1, \ldots, \xi_{M_{\max}} \}$ be an arbitrary space of dimension $M$. We assume that the space $V_M$ is orthogonal such that $ (v_n,v_m) = \delta_{nm}, 1 \le n,m \le M$ ($(\cdot,\cdot)$ denotes an appropriate inner product and $\delta_{nm}$ is the Kronecker delta symbol). The POD space, $W_M = {\rm span} \{\zeta_1, \ldots, \zeta_M\} $ is defined as

\begin{equation}\label{eq:RB_three}
W_M = \arg \min_{V_M \subset \textrm{span} \{\xi_1,\ldots,\xi_{M_{\max}}\}} \left( \displaystyle \frac{1}{M_{\max}} \displaystyle \sum_{k=1}^{M_{\max}} \inf_{\underline{\alpha}^k \in \mathbb{R}^M} \biggl\| \xi_k - \sum_{m=1}^M \alpha_m^k v_m \biggr\|^2 \right).
\end{equation}

In essence, the POD space $W_M$ which is extracted from the given set of snapshots $\{\xi_k\}_{k=1}^{M_{\max}}$ is the space that best approximate this given set of snapshots and can be written as $W_M={\rm POD}$ $\left( \{\xi_1,\ldots, \xi_{M_{\max}}\},M \right)$. We can construct this POD space by using the method of snapshots which is presented concisely in the Appendix of \cite{Nguyen2008}.

\subsubsection{Goal-oriented POD--Greedy algorithm}

We now discuss the POD--Greedy algorithms \cite{Haasdonk2008, Hoang2013} to construct the nested sets $S_*$ and $Y_N$ of interest. Let $\Xi_{\rm train}$ be a finite set of the parameters in $\mathcal{D}$ ($\Xi_{\rm train} \in \mathcal{D}$); and $S_*$ denote the set of greedily selected parameters in $\Xi_{\rm train}$. Initialize $S_* = \{\mu_0\}$, where $\mu_0$ is an arbitrarily chosen parameter. Let $e_{\rm proj}(\mu,t^k) = u(\mu,t^k) - {\rm proj}_{Y_N}u(\mu,t^k)$, where ${\rm proj}_{Y_N}u(\mu,t^k)$ is the $Y_N$-orthogonal projection of $u(\mu,t^k)$ into the $Y_N$ space. The standard and our proposed goal-oriented POD--Greedy algorithms are presented simultaneously in Table \ref{tab1}.

\begin{table}[h!]
\begin{center}
  {\begin{tabular}{| c | l | l |}
  \hline \rule{0pt}{3.0ex} \hspace{-2.2mm}
  (T\ref{tab1}a)   &   Set $Y^{\rm st}_N=0$   &   Set $Y^{\rm go}_N=0$   \\ [2ex]
  (T\ref{tab1}b)   &   Set $\mu^{\rm st}_{*}=\mu_{0}$   &   Set $\mu^{\rm go}_{*}=\mu_{0}$   \\ [2ex]
  (T\ref{tab1}c)   &    While $N \le N_{\max}$  &   While $N \le N_{\max}$   \\ [2ex]
  (T\ref{tab1}d)   &    \qquad $\mathcal{W}^{\rm st} = \left\{e^{\rm st}_{\rm proj}(\mu^{\rm st}_{*},t^k), \, 0 \le k \le K \right\}$;   &
    \qquad $\mathcal{W}^{\rm go}  = \left\{e^{\rm go}_{\rm proj}(\mu^{\rm go}_{*},t^k), \, 0 \le k \le K \right\}$;   \\ [2ex]
  (T\ref{tab1}e)   &     \qquad $Y^{\rm st}_{N+M} \longleftarrow Y^{\rm st}_N \bigoplus {\rm POD}(\mathcal{W}^{\rm st},M)$;   &
    \qquad $Y^{\rm go}_{N+M} \longleftarrow Y^{\rm go}_N \bigoplus {\rm POD}(\mathcal{W}^{\rm go},M)$;   \\ [2ex]
  (T\ref{tab1}f)   &     \qquad $N \longleftarrow N + M$;   &   \qquad $N \longleftarrow N + M$;  \\ [2ex]
  (T\ref{tab1}g)   &     \qquad $\mu^{\rm st}_{*}= \arg \max\limits_{\mu \in \Xi_{\rm train}} \left\{ \Delta_u(\mu)  \right\} $;   &
    \qquad $\mu^{\rm go}_{*}= \arg \max\limits_{\mu \in \Xi_{\rm train}} \left\{ \Delta_s(\mu) \right\} $;   \\ [2ex]
  (T\ref{tab1}h)   &     \qquad $S^{\rm st}_* \longleftarrow S^{\rm st}_* \bigcup \left\{ \mu^{\rm st}_* \right\} $;   &
    \qquad $S^{\rm go}_* \longleftarrow S^{\rm go}_* \bigcup \left\{ \mu^{\rm go}_* \right\} $;   \\ [2ex]
  (T\ref{tab1}i)   &       end.    &   end.\\
  \hline \rule{0pt}{6.0ex}
  (T\ref{tab1}j)   &     $\Delta_u(\mu) = \frac{ \sqrt{\sum_{k=1}^{K} \|\mathcal{R}^{\rm st} (v;\mu,t^k)\|^2_{Y'} } }{\sqrt{\sum_{k=1}^{K}\|u^{\rm st}_N(\mu,t^k)\|^2_Y }}$   &   $\Delta_s(\mu) = \left| \frac{ s^{\rm st}_{2N}(\mu) - s^{\rm go}_N(\mu) }{ s^{\rm st}_{2N}(\mu) } \right| $   \\ [4ex]
  \hline
\end{tabular}
\caption{(Left) Standard POD--Greedy sampling algorithm and (Right) our proposed goal-oriented POD--Greedy sampling algorithm.}
\label{tab1}}
\end{center}
\end{table}

In Table \ref{tab1}, the superscript ``st'' denotes the standard POD--Greedy and ``go'' denotes the goal-oriented POD--Greedy algorithms, respectively. The term $\|\mathcal{R}^{\rm st} (v;\mu,t^k)\|^2_{Y'}$, $\forall v \in Y$, $1 \le k \le K-1$, is the dual norm of the associated residual of equation \eqref{eq:RB_one}, namely,

\begin{multline}\label{eq:RB_four}
\mathcal{R}(v;\mu,t^k) = g^{eq}(t^k)f(v;\mu) \\
- \frac{1}{\Delta t^2} \left( m(u_N(\mu,t^{k+1}),v;\mu) - 2m(u_N(\mu,t^{k}),v;\mu) + m(u_N(\mu,t^{k-1}),v;\mu) \right) \\
- \frac{1}{\Delta t} \left( \frac{1}{2} c(u_N(\mu,t^{k+1}),v;\mu) - \frac{1}{2} c(u_N(\mu,t^{k-1}),v;\mu) \right) \\
- \left( \frac{1}{4} a(u_N(\mu,t^{k+1}),v;\mu) + \frac{1}{2} a(u_N(\mu,t^{k}),v;\mu) + \frac{1}{4} a(u_N(\mu,t^{k-1}),v;\mu) \right);
\end{multline}

\noindent and the subscript ``$2N$'' means usually as ``two times of $N$''.


\textbf{Remarks}

\begin{itemize}

  \item In essence, the term $\Delta_u (\mu)$ is the ratio of the dual norm of the residual to the RB solution. Thus, this term is roughly considered as an error indicator for the error in the solution (or field variable); and has been used mainly in the current standard POD--Greedy algorithm \cite{Haasdonk2008, Knezevic2011, Hoang2013}.

  \item On the contrary, the term $\Delta_s(\mu)$ is an asymptotic error for the interest output of the true output error (i.e., $s(\mu) - s_N(\mu)$) as $s(\mu) \approx s^{\rm st}_{2N}(\mu)$. Therefore, as observed from Table \ref{tab1}, the main difference between these two algorithms is that we somehow try to minimize the error indicator of the output functional ($\Delta_s (\mu)$) rather than an error indicator of the solution ($\Delta_u (\mu)$) by means of Greedy iterations. By this way, we expect to improve further the accuracy of the RB output computation; but consequently, we might lose the rapid convergent rate of the solution as in the standard POD--Greedy algorithm.

  \item In order to implement the goal-oriented POD--Greedy algorithm, we need to compute $s^{\rm st}_{2N}(\mu)$ for the term $\Delta_s(\mu)$. Therefore, we proposed implementing the standard POD--Greedy algorithm \textit{first} to build the basis $Y^{\rm st}_N$; and then the goal-oriented POD--Greedy algorithm is performed \textit{subsequently}. That means the offline stage of the goal-oriented algorithm is now more expensive than that of the standard one (as it also includes the offline stage of the standard one). However, the online computational cost of the goal-oriented algorithm is completely similar to that of the standard algorithm.


\end{itemize}

\subsection{Error estimations}\label{subsec_RB_EE}

\subsubsection{True errors}

We use the true errors to compare the performances of the standard and goal-oriented algorithms in the online computation stage. The relative true errors by the two algorithms for the solutions are defined as

\begin{equation}\label{eq:RB_five}
e^{\rm st}_{u}(\mu) = \frac{ \sum_{k=0}^{K-1} \int_{t^k}^{t^{k+1}} \|u(\mu,t) - u^{\rm st}_N(\mu,t) \|_Y dt}{ \sum_{k=0}^{K-1} \int_{t^k}^{t^{k+1}} \| u(\mu,t) \|_Y dt}, \quad e^{\rm go}_{u}(\mu) = \frac{ \sum_{k=0}^{K-1} \int_{t^k}^{t^{k+1}} \|u(\mu,t) - u^{\rm go}_N(\mu,t) \|_Y dt}{ \sum_{k=0}^{K-1} \int_{t^k}^{t^{k+1}} \| u(\mu,t) \|_Y dt};
\end{equation}

\noindent and for the outputs

\begin{subequations}\label{eq:RB_six}
\begin{equation}
e^{\rm st}_{s}(\mu) = \left| \frac{ s(\mu) - s^{\rm st}_N(\mu) }{ s(\mu) } \right| = \left| \frac{ \sum_{k=0}^{K-1} \int_{t^k}^{t^{k+1}} \ell(u(\mu,t) - u^{\rm st}_N(\mu,t)) dt }{ \sum_{k=0}^{K-1} \int_{t^k}^{t^{k+1}} \ell(u(\mu,t)) dt } \right|, \label{eq:RB_six-a}
\end{equation}
\begin{equation}
e^{\rm go}_{s}(\mu) = \left| \frac{ s(\mu) - s^{\rm go}_N(\mu) }{ s(\mu) } \right| = \left| \frac{ \sum_{k=0}^{K-1} \int_{t^k}^{t^{k+1}} \ell(u(\mu,t) - u^{\rm go}_N(\mu,t)) dt }{ \sum_{k=0}^{K-1} \int_{t^k}^{t^{k+1}} \ell(u(\mu,t)) dt } \right|. \label{eq:RB_six-b}
\end{equation}
\end{subequations}

In the above expressions, $u(\mu,t)$, $u^{\rm st}_N(\mu,t)$ and $u^{\rm go}_N(\mu,t)$ are the FE, standard RB and goal-oriented RB solutions; $s(\mu)$, $s^{\rm st}_N(\mu)$ and $s^{\rm go}_N(\mu)$ are the FE, standard RB and goal-oriented RB outputs, respectively.

\subsubsection{Asymptotic errors}

The true errors are good for comparison purposes but are not of practical uses in the online stage, where ones require fast and countless online calculations. We introduce the asymptotic errors for output computations as an alternative mean to evaluate relatively the quality of the RB approximations. Of course asymptotic error is not a rigorous upper error bound such as the \textit{a posteriori} error bounds \cite{Rozza2008, Nguyen2009, Knezevic2011}; however, there are several good reasons for using it in practice. First, the time-marching error bounds for the wave equation so far were shown to be ineffective and pessimistic due to the nature of the wave equation: exponential growing with respect to time \cite{HoangPhDThesis2012, Yano2012a, Yano2012b}. (We also note that the space-time error bounds for the wave equation are, although very promising, still not yet derived in the literature.) Second, the asymptotic error converges asymptotically to the true error and thus can approximate relatively the accuracy of the RB solutions/outputs for various choices of $\mu$. Thirdly -- most important, its computational cost is very cheap: only $N$-dependent (note $N \ll \mathcal{N}$). The asymptotic output errors are defined as follows

\begin{subequations}\label{eq:RB_seven}
\begin{equation}
\hat{e}^{\rm off}_{s}(\mu) = \left| \frac{ s^{\rm st}_{2N}(\mu) - s^{\rm go}_N(\mu) }{ s(\mu) } \right| = \left| \frac{ \sum_{k=0}^{K-1} \int_{t^k}^{t^{k+1}} \ell(u^{\rm st}_{2N}(\mu,t) - u^{\rm go}_N(\mu,t)) dt }{ \sum_{k=0}^{K-1} \int_{t^k}^{t^{k+1}} \ell(u(\mu,t)) dt } \right|, \label{eq:RB_seven-a}
\end{equation}
\begin{equation}
\hat{e}^{\rm onl}_{s}(\mu) = \left| \frac{ s^{\rm go}_{2N}(\mu) - s^{\rm go}_N(\mu) }{ s(\mu) } \right| = \left| \frac{ \sum_{k=0}^{K-1} \int_{t^k}^{t^{k+1}} \ell(u^{\rm go}_{2N}(\mu,t) - u^{\rm go}_N(\mu,t)) dt }{ \sum_{k=0}^{K-1} \int_{t^k}^{t^{k+1}} \ell(u(\mu,t)) dt } \right|. \label{eq:RB_seven-b}
\end{equation}
\end{subequations}

In addition, the effectivities corresponding to these asymptotic output errors can be defined as

\begin{subequations}\label{eq:RB_eight}
\begin{equation}
\eta^{\rm off}_{s}(\mu) = \left| \frac{ s^{\rm st}_{2N}(\mu) - s^{\rm go}_{N}(\mu) }{ s(\mu) - s^{\rm go}_{N}(\mu) } \right| = \frac{ \hat{e}^{\rm off}_s(\mu) }{ e^{\rm go}_s (\mu) },    \label{eq:RB_eight-a}
\end{equation}
\begin{equation}
\eta^{\rm onl}_{s}(\mu) = \left| \frac{ s^{\rm go}_{2N}(\mu) - s^{\rm go}_{N}(\mu) }{ s(\mu) - s^{\rm go}_{N}(\mu) } \right| = \frac{ \hat{e}^{\rm onl}_s(\mu) }{ e^{\rm go}_s (\mu) }.    \label{eq:RB_eight-b}
\end{equation}
\end{subequations}

We note that $\hat{e}^{\rm off}_{s}(\mu)$ is similar to $\Delta_s(\mu)$ -- the error indicator for the output in the offline stage; and $\hat{e}^{\rm onl}_{s}(\mu)$ is the useful asymptotic error practically used in the online stage of the goal-oriented POD--Greedy algorithm.

\subsection{Offline-online computational procedure}\label{subsec_RB_OffOn}

In this section, we develop offline-online computational procedures in order to fully exploit the dimension reduction of the problem \cite{Grepl2005, Hoang2013, Nguyen2009}. We note that both algorithms (standard and goal-oriented) have the same offline-online computational procedures, they are only different in the ways to build the sets $S_*$ and $Y_N$ within Greedy iterations. We first express $u_N(\mu,t^k)$ as:

\begin{equation}\label{eq:RB_nine}
u_N(\mu,t^k) = \sum_{n=1}^{N} u_{N \, n}(\mu,t^k) \; \zeta_n,   \qquad \forall \zeta_n \in Y_N.
\end{equation}

We then choose a test function $v = \zeta_n, 1 \le n \le N$ for the RB equation \eqref{eq:RB_one}. It then follows that $\underline{u}_N(\mu,t^k) = [u_{N \, 1}(\mu,t^k) \; u_{N \, 2}(\mu,t^k) \; \cdots \; u_{N \, N}(\mu,t^k) ]^T \in \mathbb{R}^N$ satisfies

\begin{multline}\label{eq:RB_ten}
\left( \frac{1}{\Delta t^2} \mathbf{M}_N(\mu) + \frac{1}{2 \Delta t} \mathbf{C}_N(\mu) + \frac{1}{4} \mathbf{A}_N(\mu) \right) \underline{u}_N(\mu,t^{k+1}) \\
= \left( - \frac{1}{\Delta t^2} \mathbf{M}_N(\mu) + \frac{1}{2 \Delta t} \mathbf{C}_N(\mu) - \frac{1}{4} \mathbf{A}_N(\mu) \right) \underline{u}_N(\mu,t^{k-1}) \\
+ \left( \frac{2}{\Delta t^2} \mathbf{M}_N(\mu) - \frac{1}{2} \mathbf{A}_N(\mu) \right) + g^{eq}(t^k) \mathbf{F}_N(\mu), \qquad 1 \leq k \leq K-1.
\end{multline}

The initial condition is treated similar to the treatment in \eqref{eq:ProbStat_six} and \eqref{eq:RB_one}. Here, $\mathbf{C}_N(\mu)$, $\mathbf{A}_N(\mu)$, $\mathbf{M}_N(\mu) \in \mathbb{R}^{N \times N}$ are symmetric positive definite matrices with entries $\mathbf{C}_{N \, i,j}(\mu) = c(\zeta_i,\zeta_j;\mu)$, $\mathbf{A}_{N \, i,j}(\mu) = a(\zeta_i,\zeta_j;\mu)$, $\mathbf{M}_{N \, i,j}(\mu) = m(\zeta_i,\zeta_j;\mu)$, $1 \le i,j \le N$ and $\mathbf{F}_N \in \mathbb{R}^N$ is the RB load vector with entries $\mathbf{F}_{N \, i} = f(\zeta_i)$, $1 \le i \le N$, respectively.

The RB output is then computed from

\begin{equation}\label{eq:RB_eleven}
s_N(\mu) = \sum_{k=0}^{K-1} \int_{t^k}^{t^{k+1}} \mathbf{L}^T_N \, \underline{u}_N(\mu,t) dt.
\end{equation}

Invoking the affine parameter dependence \eqref{eq:ProbStat_ten}, we obtain

\begin{subequations}\label{eq:RB_twelve}
\begin{equation}
\mathbf{M}_{N \, i,j}(\mu) = m(\zeta_i,\zeta_j;\mu) = \sum_{q=1}^{Q_m} \Theta^q_m(\mu) \, m^q(\zeta_i,\zeta_j), \label{eq:RB_twelve-a}
\end{equation}
\begin{equation}
\mathbf{C}_{N \, i,j}(\mu) = c(\zeta_i,\zeta_j;\mu) = \sum_{q=1}^{Q_c} \Theta^q_c(\mu) \, c^q(\zeta_i,\zeta_j), \label{eq:RB_twelve-b}
\end{equation}
\begin{equation}
\mathbf{A}_{N \, i,j}(\mu) = a(\zeta_i,\zeta_j;\mu) = \sum_{q=1}^{Q_a} \Theta^q_a(\mu) \, a^q(\zeta_i,\zeta_j), \label{eq:RB_twelve-c}
\end{equation}
\begin{equation}
\mathbf{F}_{N \, i}(\mu) = f(\zeta_i;\mu) = \sum_{q=1}^{Q_f} \Theta^q_f(\mu) \, f^q(\zeta_i), \label{eq:RB_twelve-d}
\end{equation}
\end{subequations}

\noindent which can be written as

\begin{align}\label{eq:RB_thirteen}
\mathbf{M}_{N \, i,j}(\mu) = \sum_{q=1}^{Q_m} \Theta^q_m(\mu) \, \mathbf{M}_{N \, i,j}^q, &\qquad  \mathbf{C}_{N \, i,j}(\mu) = \sum_{q=1}^{Q_c} \Theta^q_c(\mu) \, \mathbf{C}_{N \, i,j}^q,  \nonumber \\
\mathbf{A}_{N \, i,j}(\mu) = \sum_{q=1}^{Q_a} \Theta^q_a(\mu) \, \mathbf{A}_{N \, i,j}^q, &\qquad  \mathbf{F}_{N \, i}(\mu) = \sum_{q=1}^{Q_f} \Theta^q_f(\mu) \, \mathbf{F}_{N \, i}^q,
\end{align}

\noindent where the \textit{parameter independent} quantities $\mathbf{M}_N^q$, $\mathbf{C}_N^q$, $\mathbf{A}_N^q \in \mathbb{R}^{N \times N}$, and $\mathbf{F}_N^q \in \mathbb{R}^N$ are given by

\begin{subequations}\label{eq:RB_fourteen}
\begin{equation}
\mathbf{M}_{N \, i,j}^q = m^q(\zeta_i,\zeta_j), \quad 1 \le i,j \le N_{\max}, \quad 1 \le q \le Q_m, \label{eq:RB_fourteen-a}
\end{equation}
\begin{equation}
\mathbf{C}_{N \, i,j}^q = c^q(\zeta_i,\zeta_j), \quad 1 \le i,j \le N_{\max}, \quad 1 \le q \le Q_c, \label{eq:RB_fourteen-b}
\end{equation}
\begin{equation}
\mathbf{A}_{N \, i,j}^q = a^q(\zeta_i,\zeta_j), \quad 1 \le i,j \le N_{\max}, \quad 1 \le q \le Q_a, \label{eq:RB_fourteen-c}
\end{equation}
\begin{equation}
\mathbf{F}_{N \, i}^q = f^q(\zeta_i), \quad 1 \le i \le N_{\max}, \quad 1 \le q \le Q_f, \label{eq:RB_fourteen-d}
\end{equation}
\end{subequations}

\noindent respectively.

The offline-online computational procedure is now described as follows. In the offline stage -- performed only once, we first implement the standard POD--Greedy algorithm \cite{Hoang2013}: we solve for the $\zeta^{\rm st}_n, 1 \le n \le N_{\max}$; then compute and store the $\mu$-independent quantities in \eqref{eq:RB_fourteen} for the estimation of the RB solution and output\footnotemark \footnotetext{There are still several terms related to the computation of dual norm of the residual, we do not show them here for simplicity. We refer the readers to \cite{Hoang2013} for all related details.}. Once these RB solution and output are available, we can now implement the goal-oriented POD--Greedy algorithm. We consider each goal-oriented POD--Greedy iteration (Table \ref{tab1}) in more details. We first need to solve \eqref{eq:ProbStat_six} for the ``true'' FE solutions; then do the error projection in step (T\ref{tab1}d) and solve the POD/eigenvalue problem in step (T\ref{tab1}e). In addition, we have to compute $O(N^2 Q)$ $\mathcal{N}$-inner products in \eqref{eq:RB_fourteen} (we denote $Q=Q_m+Q_c+Q_a$). By approximating $s(\mu)$ by the enriched approximation $s^{\rm st}_{2N}(\mu)$ in (T\ref{tab1}j) through the standard algorithm, we can now do a cheaply exhaustive/wide search over $\Xi_{\rm train}$ to look for optimal $\mu$ in each Greedy iteration. In summary, the offline stage of the goal-oriented algorithm also includes the offline stage of the standard algorithm, and therefore it is more expensive than that of the standard one.

The online stage of the goal-oriented algorithm is completely similar to that of the standard algorithm \cite{Hoang2013}. In the online stage -- performed many times, for each new parameter $\mu$ -- we first assemble the RB matrices in \eqref{eq:RB_twelve}, this requires $O(N^2 Q)$ operations. We then solve the RB governing equation \eqref{eq:RB_ten}, the operation counts are $O(N^3+KN^2)$ as the RB matrices are generally full. Finally, we evaluate the displacement output $s_N(\mu)$ from \eqref{eq:RB_eleven} at the cost of $O(KN)$. For the online asymptotic error (i.e., $s^{\rm go}_{2N}(\mu) - s^{\rm go}_{N}(\mu)$), there is nothing more than computing one more output $s^{\rm go}_{2N}(\mu)$ and then performing the associated subtraction; the cost is $O(2KN)$. Therefore, as required in real-time context, the online complexity to compute the output and its associated asymptotic error are independent of $\mathcal{N}$, and since $N \ll \mathcal{N}$ we can expect significant computational saving in the online stage relative to the classical FE approach.

\section{Numerical example}\label{sec_NumEx}

In this section, we will verify both POD--Greedy algorithms by investigating an numerical example which is a three-dimensional dental implant model problem in the time domain. This model problem is similar to that in the work of Hoang et al. \cite{Hoang2013}. The details are described in the following.

\subsection{A 3D dental implant model problem}\label{subsec_NumEx_Model}

\begin{figure}[h!]
 \centering
 \subfigure[]{ \includegraphics[height=5.5cm]{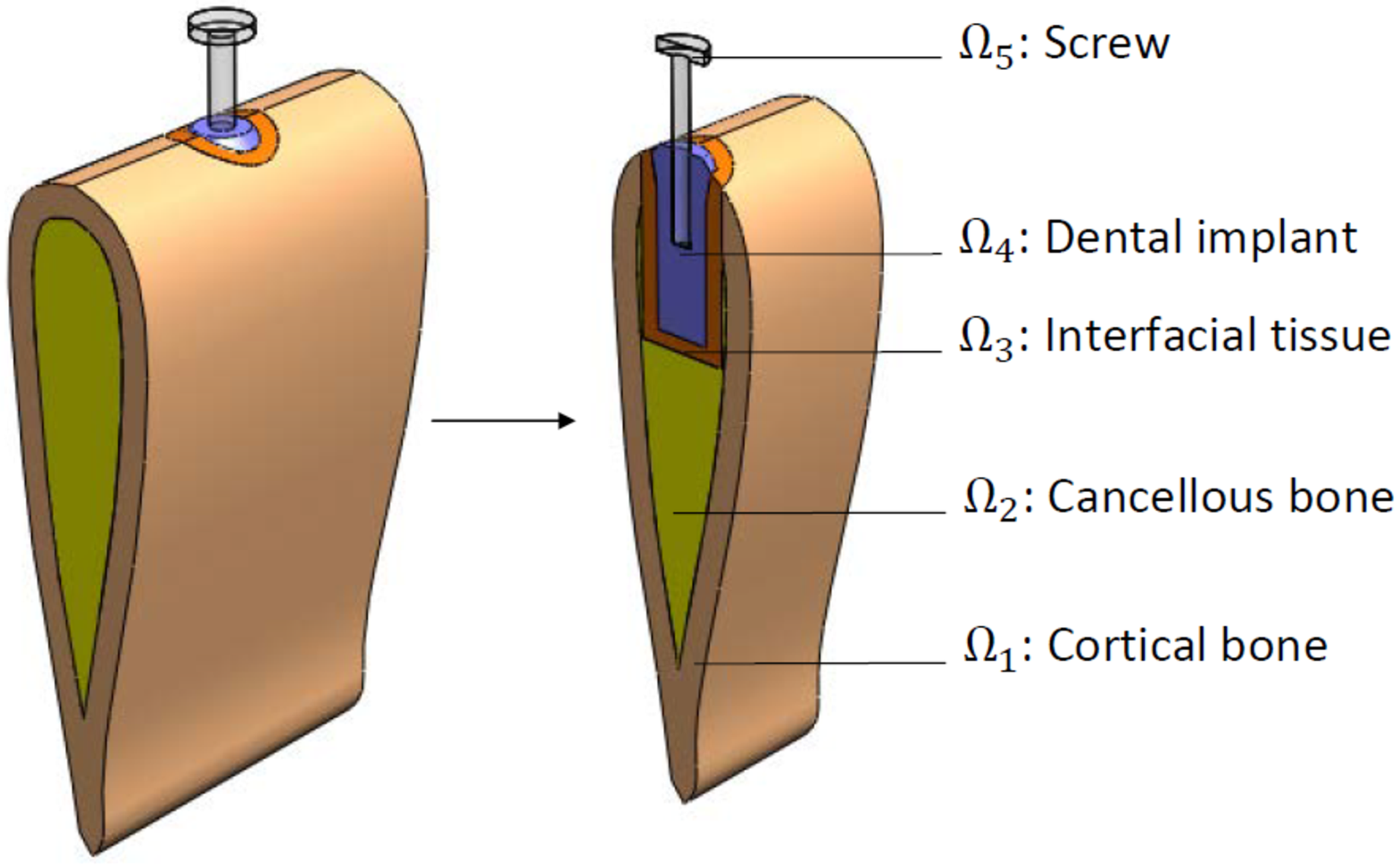}
 \label{fig1a}}
 \subfigure[]{ \includegraphics[height=6.3cm]{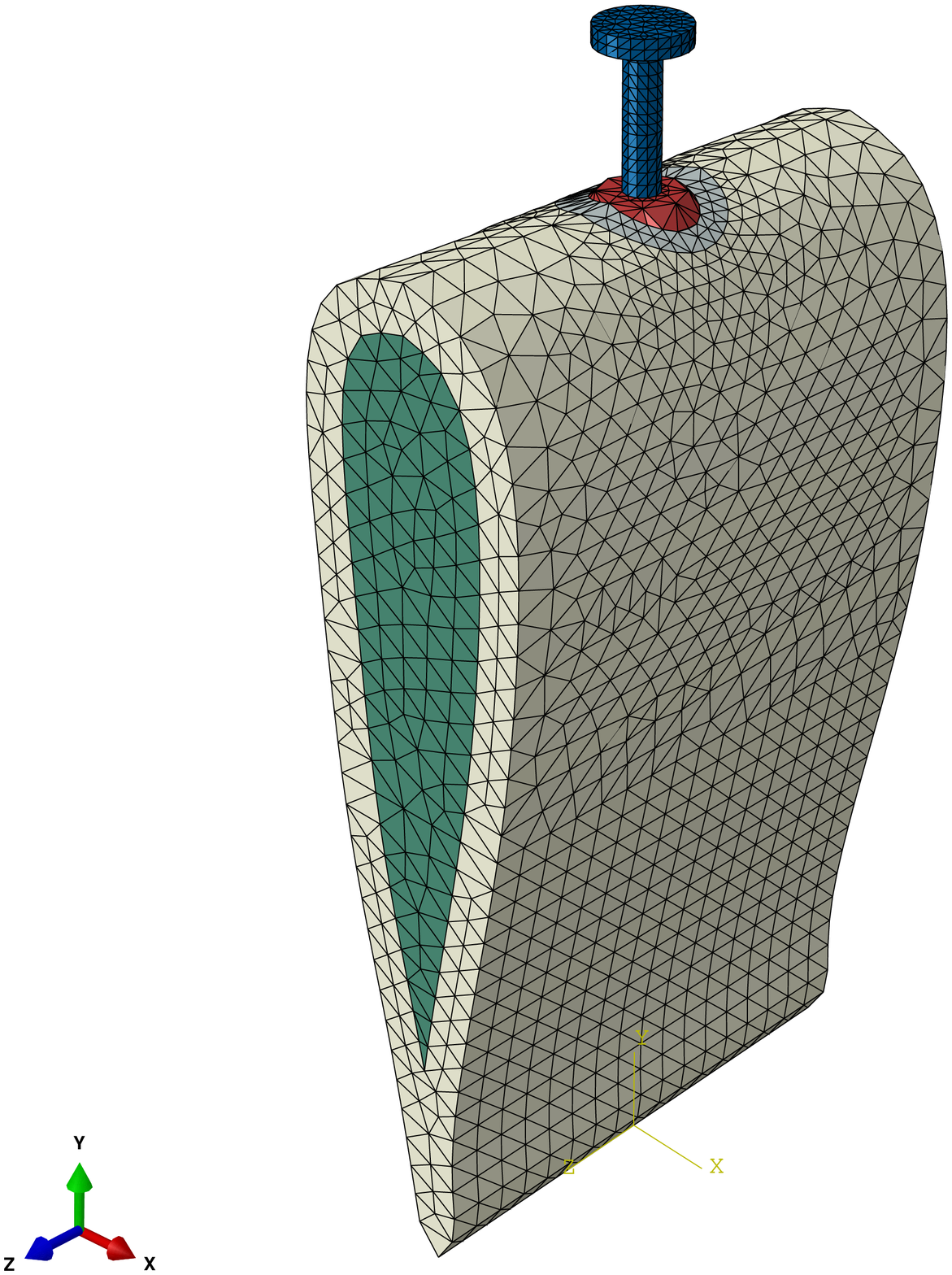}
 \label{fig1b}}
 \caption{(a) The 3D simplified FEM model with sectional view, and (b) meshing in ABAQUS.}
 \label{fig1}
\end{figure}

\begin{figure}[h!]
 \centering
 \subfigure[]{ \includegraphics[height=6cm]{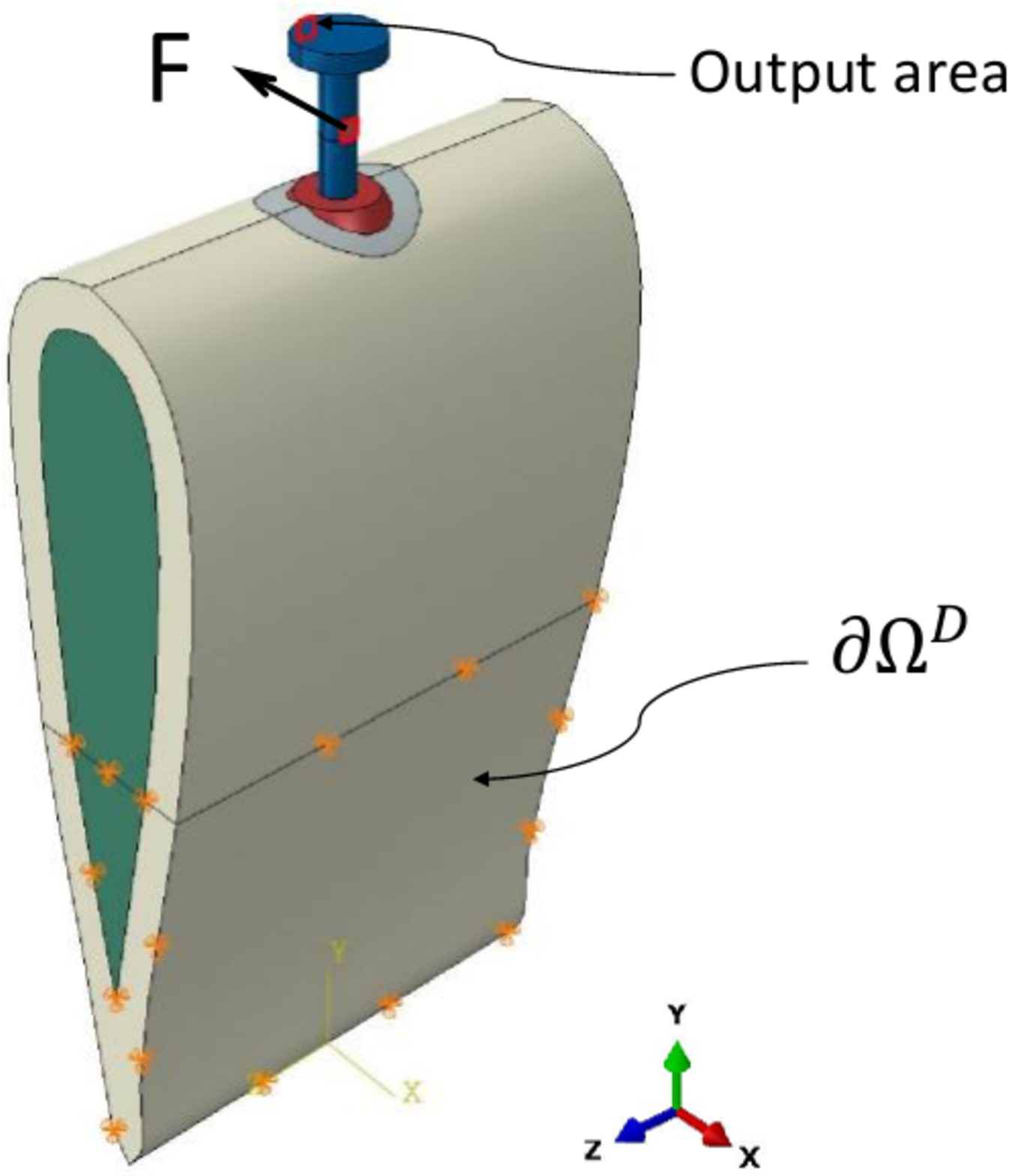}
 \label{fig2a}}
 \subfigure[]{ \includegraphics[height=6cm]{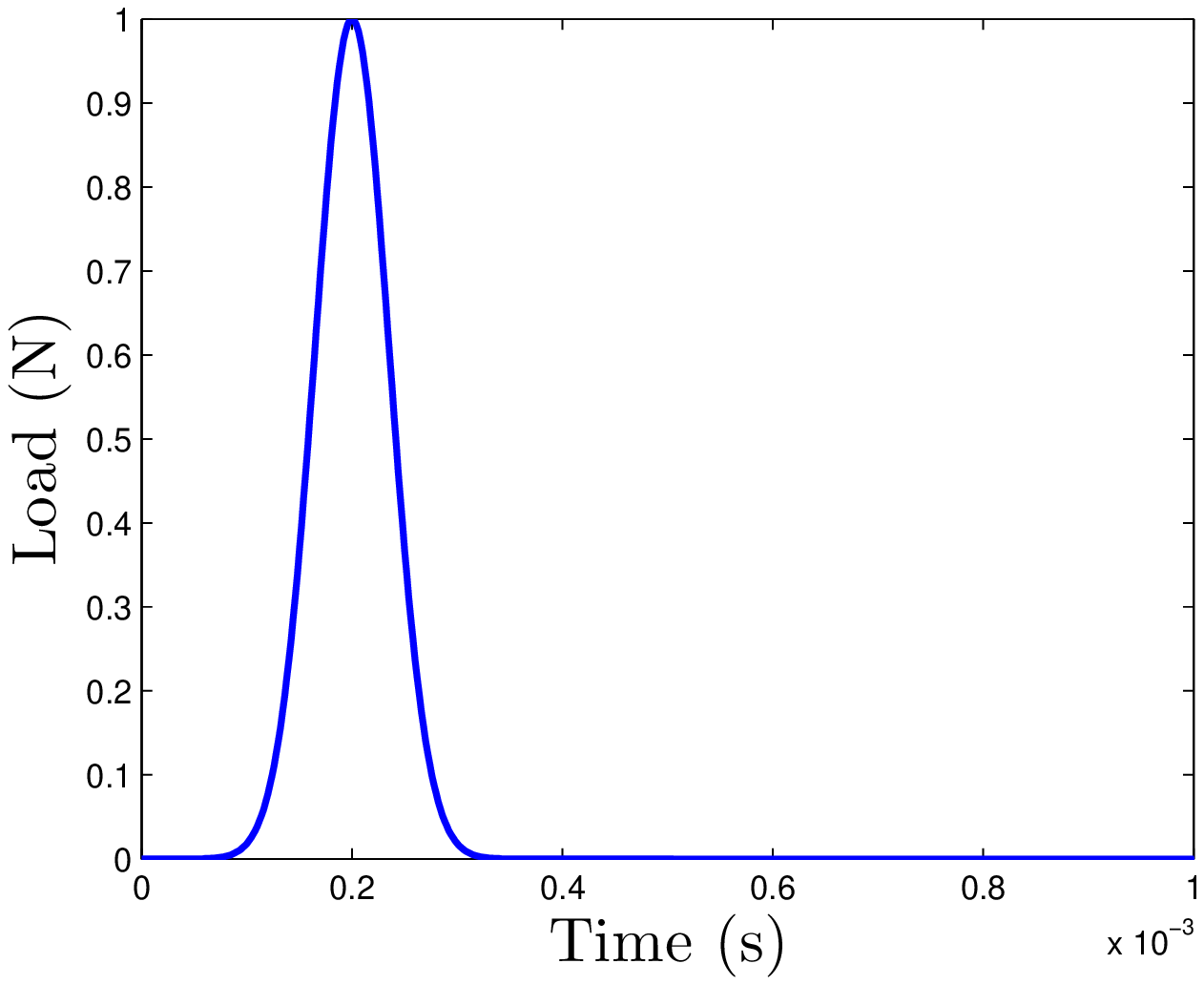}
 \label{fig2b}}
 \caption{(a) Output area, applied load {\bf F} and boundary condition, and (b) time history of a particular load.}
 \label{fig2}
\end{figure}

We consider a simplified 3D dental implant-bone model in Fig.\ref{fig1a}. The geometry of the simplified dental implant-bone model is constructed by using SolidWorks 2010. The physical domain $\Omega$ consists of five regions: the outermost cortical bone $\Omega_1$, the cancellous bone $\Omega_2$, the interfacial tissue $\Omega_3$, the dental implant $\Omega_4$ and the stainless steel screw $\Omega_5$. The 3D simplified model is then meshed and analyzed in the software ABAQUS/CAE version 6.10-1 (Fig.\ref{fig1b}). A dynamic force opposite to the $x-$direction is then applied to a prescribed area on the body of the screw as shown in Fig.\ref{fig2a}. As mentioned in Section \ref{subsec_ProbStat_ImpRes}, all computations and simulations will be performed for the unit input loading case, since the solutions associated with other input loading cases can be easily inferred from the Duhamel's convolution principle \eqref{eq:ProbStat_eleven}. For example, Fig.\ref{fig2b} shows the time history of such ``arbitrary'' input loading case. The output of interest is defined as the average displacement responses of a prescribed area on the head of the screw (Fig.\ref{fig2a}). The Dirichlet boundary condition $(\partial \Omega^D)$ is specified in the bottom-half of the simplified model as illustrated in Fig.\ref{fig2a}. The finite element mesh consists of 9479 nodes and 50388 four-node tetrahedral solid elements. The coinciding nodes of the contact surfaces between different regions (the regions $\Omega_1$, $\Omega_2$, $\Omega_3$, $\Omega_4$, $\Omega_5$) are assumed to be rigidly fixed, i.e. the displacements in the $x-$, $y-$ and $z-$directions are all set to be the same for the same coinciding nodes.

We assume that the regions $\Omega_i, 1 \leq i \le 5$, of the simplified model are homogeneous and isotropic. The material properties: the Young's moduli, Poisson's ratios and densities of these regions are presented in Table \ref{tab2} \cite{Wang2010}. As similar to \cite{Hoang2013}, we still use Rayleigh damping with stiffness-proportional damping coefficient $\beta_i$, $1 \le i \le 5$ (Table \ref{tab2}) such that $\textbf{C}_i = \beta_i \textbf{A}_i, \, 1 \le i \le 5$, where $\textbf{C}_i$ and $\textbf{A}_i$ are the FEM damping and stiffness matrices of each region, respectively. We also note in Table \ref{tab2} that ($E$,$\beta$) -- the Young's modulus and Rayleigh damping coefficient associated with the interfacial tissue are our sole parameters.

\begin{table}[h!]
\begin{center}
  {\begin{tabular}{|c|l|r|l|r|r|}
  \hline \rule{0pt}{2.5ex}
 Domain & Layers & E (Pa) & $\nu$ & $\rho (\rm g/mm^3)$ & $\beta$ \\ [0.5ex]
  \hline \rule{0pt}{2.5ex}
 $\Omega_1$ & Cortical bone         & $2.3162 \times 10^{10}$ & 0.371  & $1.8601 \times 10^{-3}$ & $3.38 \times 10^{-6}$ \\
 $\Omega_2$ & Cancellous bone       & $8.2345 \times 10^{8}$  & 0.3136 & $7.1195 \times 10^{-4}$ & $6.76 \times 10^{-6}$ \\
 $\Omega_3$ &  Tissue                &      E                  & 0.3155 & $1.055  \times 10^{-3}$ &         $\beta$ \\
 $\Omega_4$ &  Titan implant         & $1.05  \times 10^{11}$  & 0.32   & $4.52   \times 10^{-3}$ & $5.1791 \times 10^{-10}$ \\
 $\Omega_5$ &  Stainless steel screw & $1.93  \times 10^{11}$  & 0.305  & $8.027  \times 10^{-3}$ & $2.5685 \times 10^{-8}$ \\ [0.3ex]
   \hline
\end{tabular}
\caption{Material properties of the dental implant-bone structure.}
\label{tab2}}
\end{center}
\end{table}

Finally, with respect to our particular dental implant problem, the actual integral forms of the linear and bilinear forms are defined as:

\begin{subequations}\label{eq:NumEx_one}
\begin{equation}
m(w,v) = \sum_{r=1}^{5} \int_{\Omega_r} \rho_r w_i v_i, \label{eq:NumEx_one-a}
\end{equation}
\begin{equation}
a(w,v;\mu) = \sum_{r=1, r \neq 3}^{5} \int_{\Omega_r} \frac{\partial v_i}{\partial x_j} C_{ijkl}^{r} \frac{\partial w_k}{\partial x_l} + \mu_{1} \int_{\Omega_3} \frac{\partial v_i}{\partial x_j} C_{ijkl}^3 \frac{\partial w_k}{\partial x_l}, \label{eq:NumEx_one-b}
\end{equation}
\begin{equation}
c(w,v;\mu) = \sum_{r=1, r \neq 3}^{5} \beta_r \int_{\Omega_r} \frac{\partial v_i}{\partial x_j} C_{ijkl}^r \frac{\partial w_k}{\partial x_l} + \mu_{2}\mu_{1} \int_{\Omega_3} \frac{\partial v_i}{\partial x_j} C_{ijkl}^3 \frac{\partial w_k}{\partial x_l}, \label{eq:NumEx_one-c}
\end{equation}
\begin{equation}
f(v) = \int_{\Gamma_{\rm l}} v, \label{eq:NumEx_one-d}
\end{equation}
\begin{equation}
\ell(v) = \frac{1}{|\Gamma_{\rm o}|} \int_{\Gamma_{\rm o}} v, \label{eq:NumEx_one-e}
\end{equation}
\end{subequations}

\noindent for all $w,v \in Y$, $\mu \in \mathcal{D}$. Here, the parameter $\mu=(\mu_1, \mu_2) \equiv (E,\beta)$ belongs to the region $\Omega_3$. $C_{ijkl}^r$ is the constitutive elasticity tensor for isotropic materials and it is expressed in terms of the Young's modulus $E$ and Poisson's ratio $\nu$ of each region $\Omega_r, 1 \leq r \leq 5$, respectively. $\Gamma_{\rm l}$ is the prescribed loading area (surface traction) and $\Gamma_{\rm o}$ is the prescribed output area as shown in Fig.\ref{fig2a}, respectively. It is obvious from \eqref{eq:ProbStat_ten} and \eqref{eq:NumEx_one} that the smooth functions $\Theta^1_a(\mu) = 1$, $\Theta^2_a(\mu) = \mu_1$; $\Theta^1_c(\mu) = 1$, $\Theta^2_c(\mu) = \mu_1 \mu_2$ depend on $\mu$ --- but the bilinear forms $a^1(w,v) = \sum_{r=1, r \neq 3}^{5} \int_{\Omega_r} \frac{\partial v_i}{\partial x_j} C_{ijkl}^{r} \frac{\partial w_k}{\partial x_l}$, $a^2(w,v) = \int_{\Omega_3} \frac{\partial v_i}{\partial x_j} C_{ijkl}^3 \frac{\partial w_k}{\partial x_l}$; $c^1(w,v) = \sum_{r=1, r \neq 3}^{5} \beta_r \int_{\Omega_r} \frac{\partial v_i}{\partial x_j} C_{ijkl}^r \frac{\partial w_k}{\partial x_l}$ and $c^2(w,v) = \int_{\Omega_3} \frac{\partial v_i}{\partial x_j} C_{ijkl}^3 \frac{\partial w_k}{\partial x_l}$ do \textit{not} depend on $\mu$.

\subsection{Numerical results}\label{subsec_NumEx_NumRes}

\begin{figure}[h!]
\begin{center}
\subfigure[]{
\resizebox*{7cm}{!}{\includegraphics{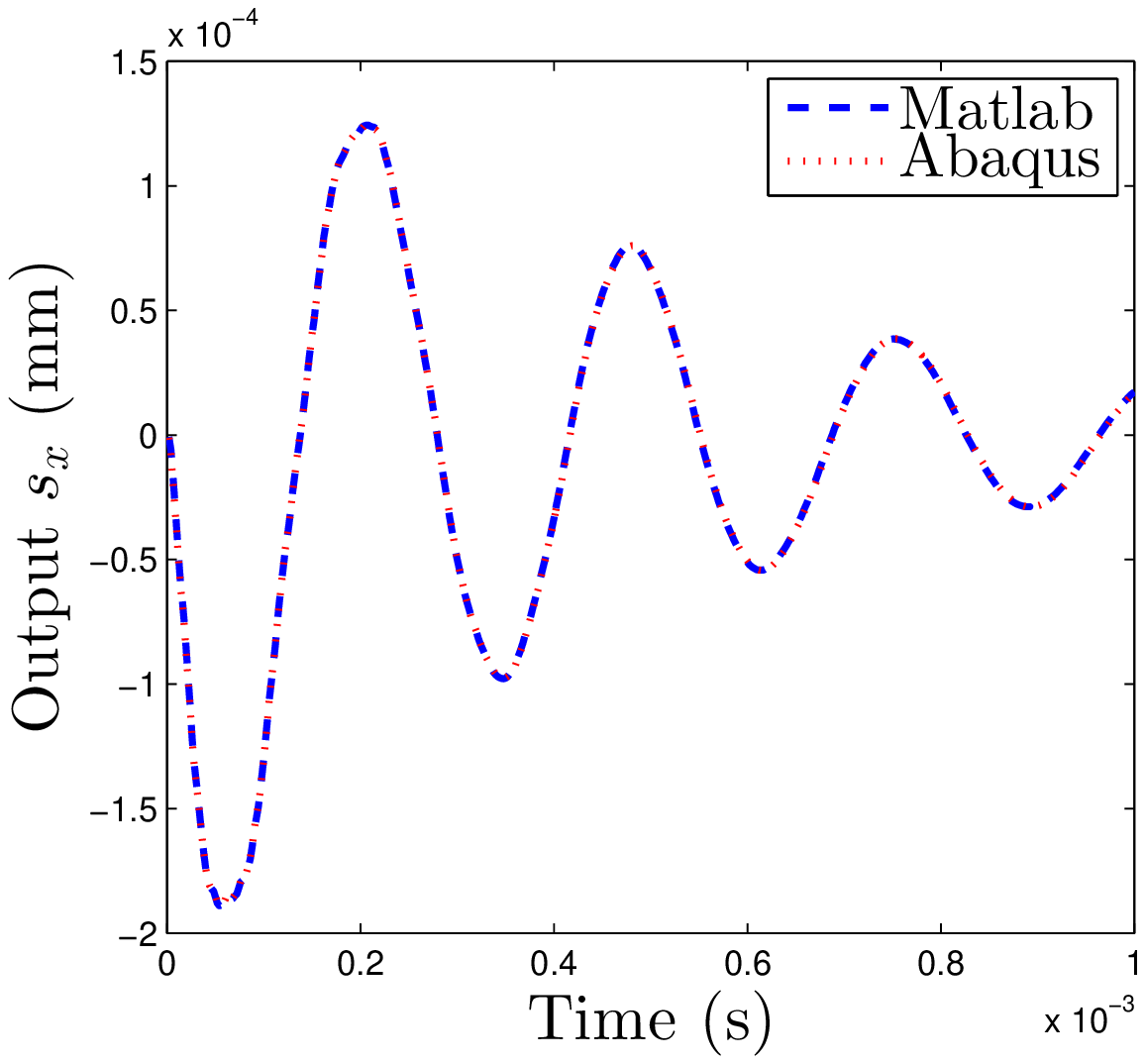}} \label{fig3a} }\\%
\subfigure[]{
\resizebox*{7cm}{!}{\includegraphics{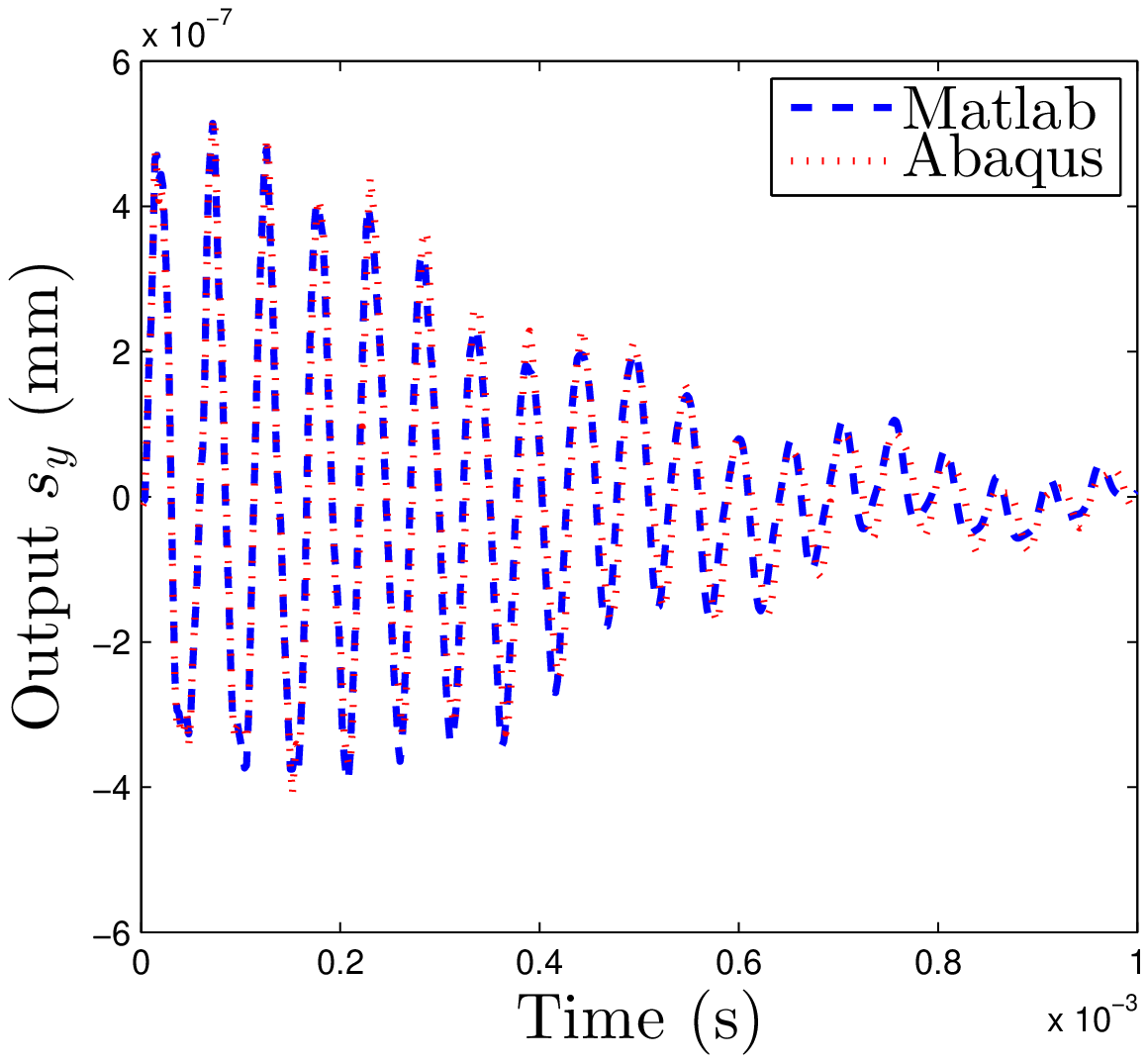}} \label{fig3b} }%
\subfigure[]{
\resizebox*{7cm}{!}{\includegraphics{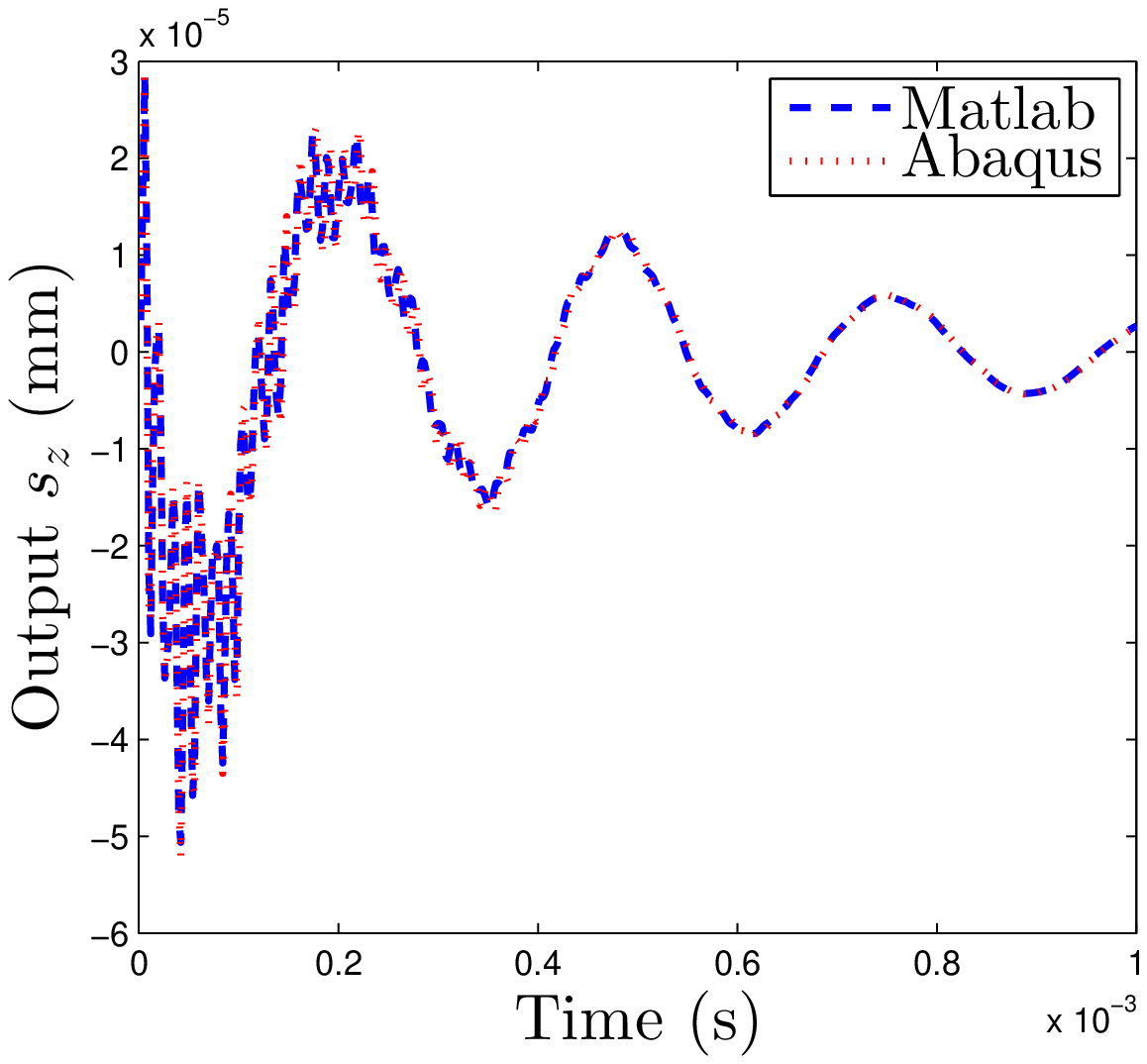}} \label{fig3c} }%
\caption{Comparison of the ``unit'' FEM output displacements computed by our code versus by ABAQUS software with respect time in the $x-$ (a), $y-$ (b), and $z-$direction (c) with $\mu_{\rm test}=(10 \times 10^6{\rm Pa}, 1 \times 10^{-5})$.}%
\label{fig3}
\end{center}
\end{figure}

\begin{figure}[h!]
\begin{center}
\subfigure[]{
\resizebox*{7cm}{!}{\includegraphics{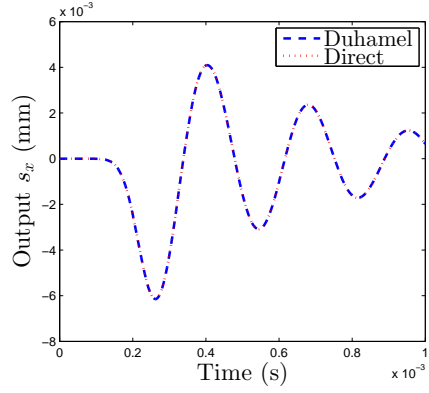}} \label{fig4a} }\\%
\subfigure[]{
\resizebox*{7cm}{!}{\includegraphics{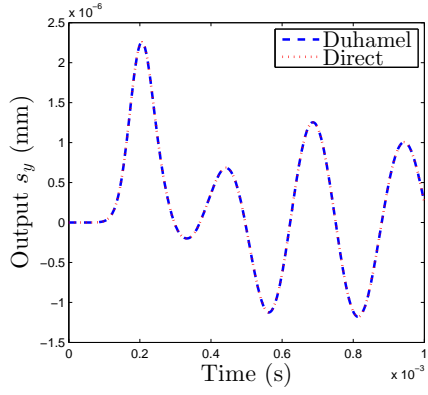}} \label{fig4b} }%
\subfigure[]{
\resizebox*{7cm}{!}{\includegraphics{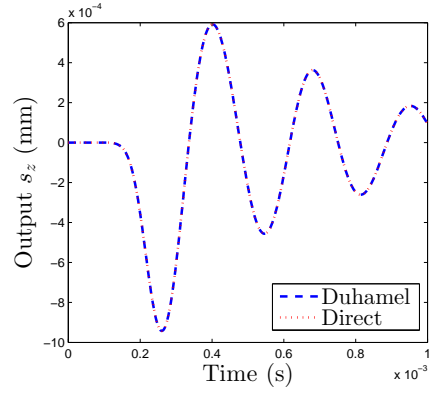}} \label{fig4c} }%
\caption{Comparison of ``arbitrary'' FEM output displacements (with the applied load in Fig.\ref{fig2b}) computed by Duhamel's convolution and direct computation with respect time in the $x-$ (a), $y-$ (b), and $z-$direction (c) with $\mu_{\rm test}=(10 \times 10^6{\rm Pa}, 1 \times 10^{-5})$.}%
\label{fig4}
\end{center}
\end{figure}

\begin{figure}[h!]
\begin{center}
\resizebox*{7cm}{!}{\includegraphics{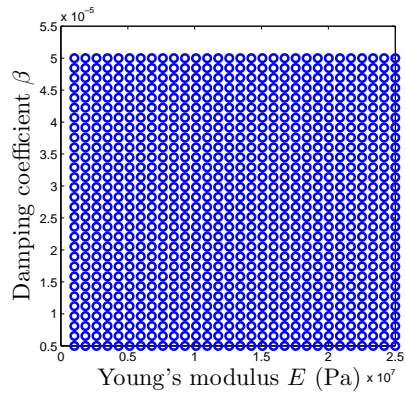}}
\caption{The training set $\Xi_{\rm train}$ with $n_{\rm train}=900$ samples.}
\label{fig5}
\end{center}
\end{figure}

The FE space to approximate the 3D dental implant-bone problem is of dimension $\mathcal{N}=26343$. For time integration, $T=1 \times 10^{-3}$s, $\Delta t = 2 \times 10^{-6}$s, $K=\frac{T}{\Delta t}=500$. The input parameter $\mu \equiv (E,\beta) \in \mathcal{D}$, where the parameter domain $\mathcal{D} \equiv [1 \times 10^6, 25 \times 10^6]{\rm Pa} \times [5 \times 10^{-6}, 5 \times 10^{-5}] \subset \mathbb{R}^{P=2}$. (Note that the range of $E$ of this parameter domain is nearly two times larger than that of \cite{Hoang2013}.) The $\|\cdot\|_{Y}$ norm used in this work is defined as $\|w\|_Y^2 = a(w,w;\bar{\mu}) + m(w,w;\bar{\mu})$, where $\bar{\mu} = (13 \times 10^6 {\rm Pa}, 2.75 \times 10^{-5})$ is the arithmetic average of $\mu$ in $\mathcal{D}$; $Q_a=2$, $Q_c=2$. To verify our computational code (performed in Matlab R2012b), we first compare the FEM outputs computed by ABAQUS and by our code with the test parameter $\mu_{\rm test}=(10 \times 10^6{\rm Pa}, 1 \times 10^{-5})$. Fig.\ref{fig3} shows the ``unit'' output displacements (i.e., under the unit loading case) in the $x-$, $y-$ and $z-$directions versus time at $\mu_{\rm test}$ via ABAQUS and our code, respectively. Fig.\ref{fig3} demonstrates that the FEM results by our code match very well with the results computed by ABAQUS. Next, we show in Fig.\ref{fig4} the FEM output displacements versus time under the loading history in Fig.\ref{fig2b} by direct FEM computation and by Duhamel's convolution. It is observed that these two results match perfectly well with each other.

We now discuss the POD--Greedy algorithms of interest. As shown in Fig.\ref{fig5}, a sample set $\Xi_{\rm train}$ is created by a uniform distribution over $\mathcal{D}$ with $n_{\rm train}(=30 \times 30) = 900$ samples. Note that we use $M=1$ and $N_{\max}=120$ (as in Table \ref{tab1}) to terminate the iteration procedures.

\begin{figure}[h!]
\begin{center}
\subfigure[]{
\resizebox*{7cm}{!}{\includegraphics{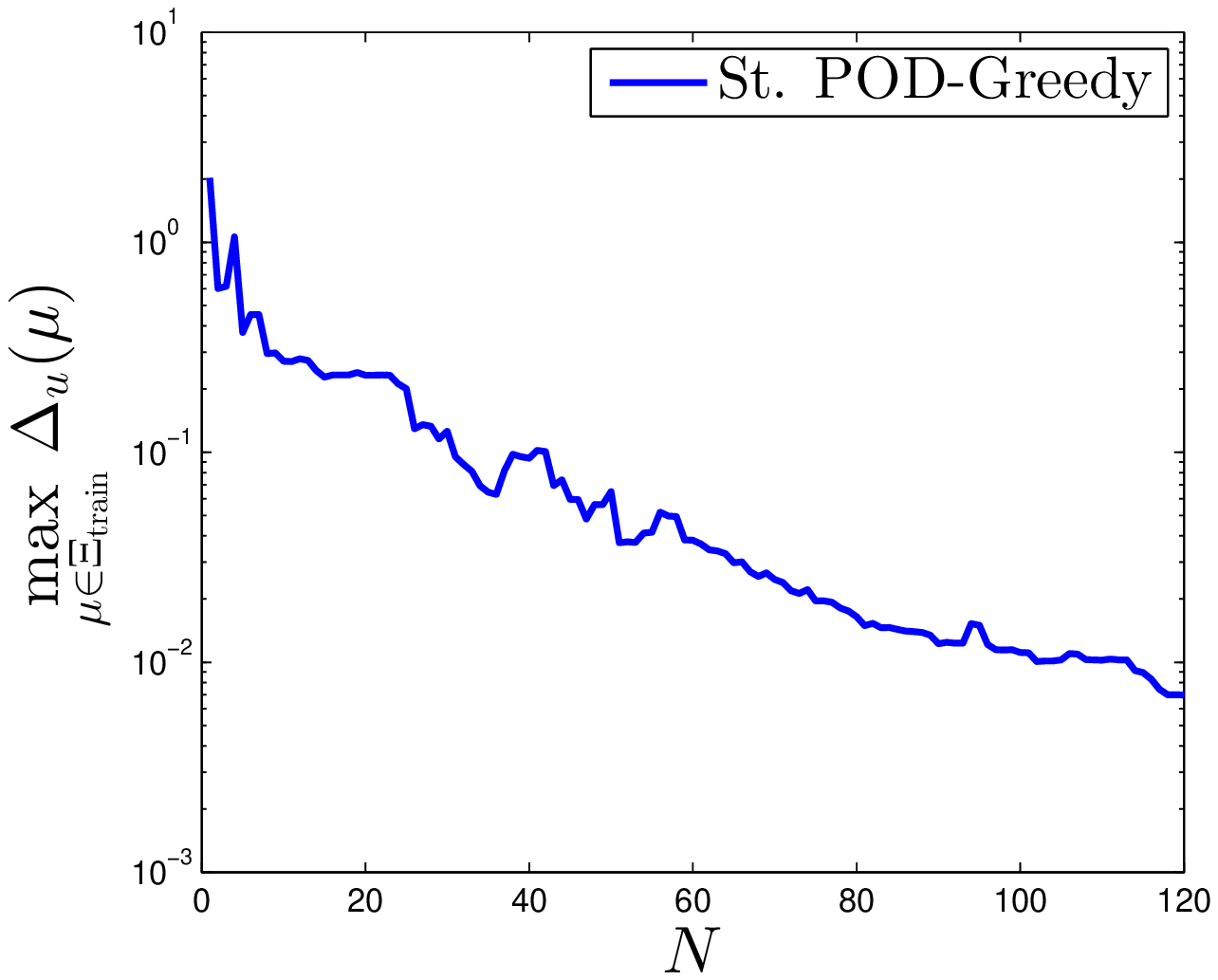}} \label{fig6a} }%
\subfigure[]{
\resizebox*{7cm}{!}{\includegraphics{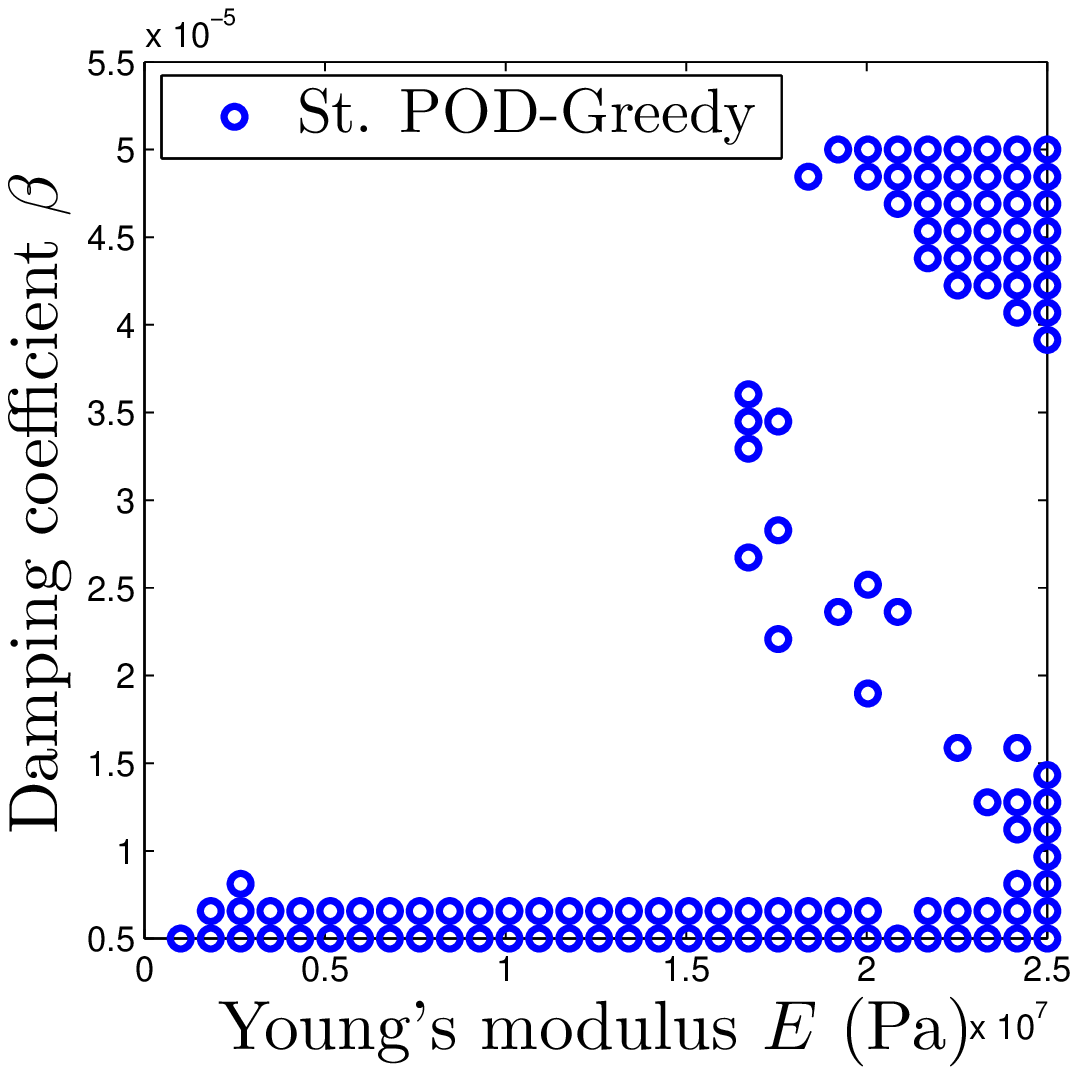}} \label{fig6b} }%
\caption{(a) Maximum of error indicator $\Delta_u(\mu)$ over $\Xi_{\rm train}$ as a function of $N$, and (b) distribution of sampling points by the standard POD--Greedy algorithm.}%
\label{fig6}
\end{center}
\end{figure}

\begin{figure}[h!]
\begin{center}
\subfigure[]{
\resizebox*{7cm}{!}{\includegraphics{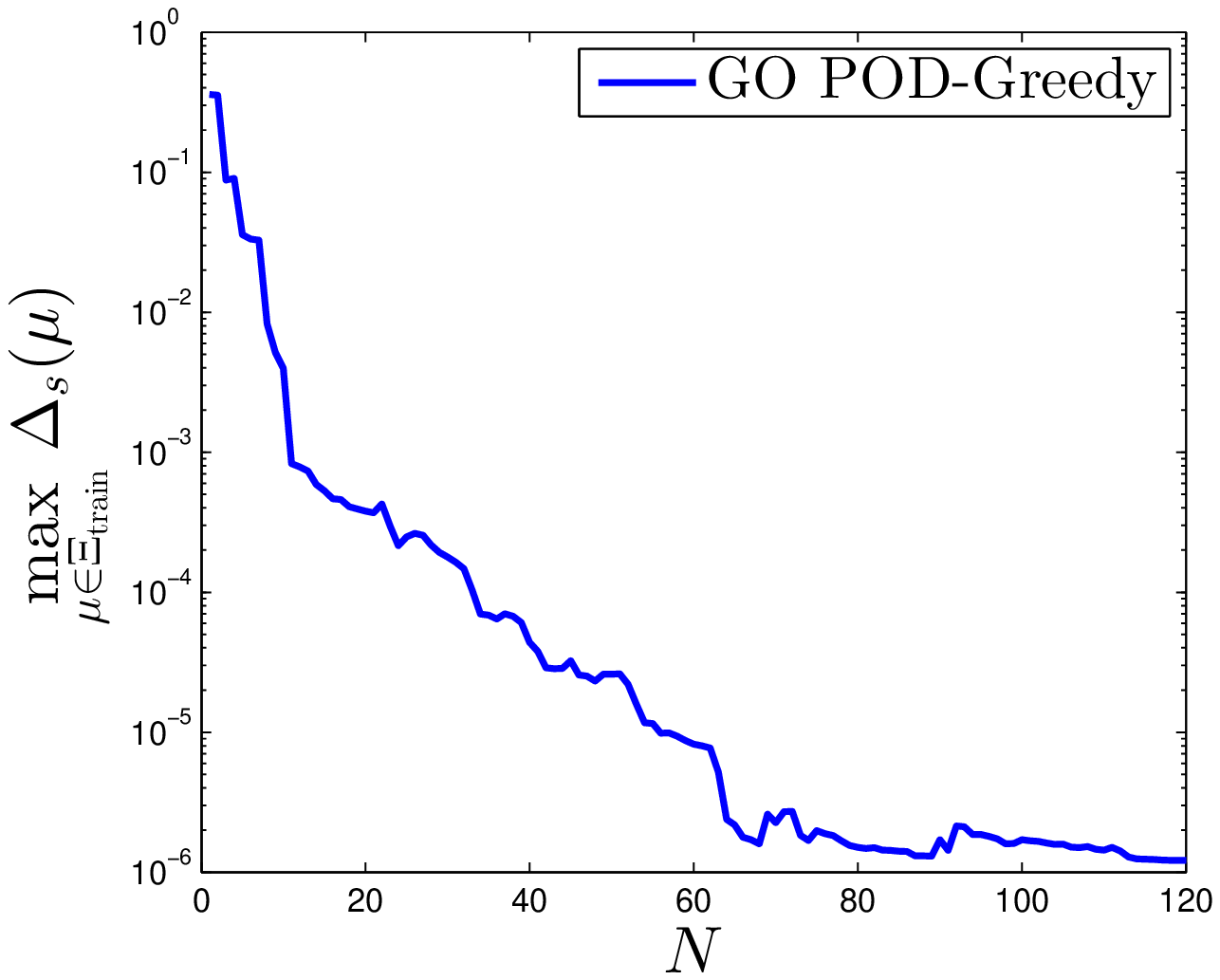}} \label{fig7a} }%
\subfigure[]{
\resizebox*{7cm}{!}{\includegraphics{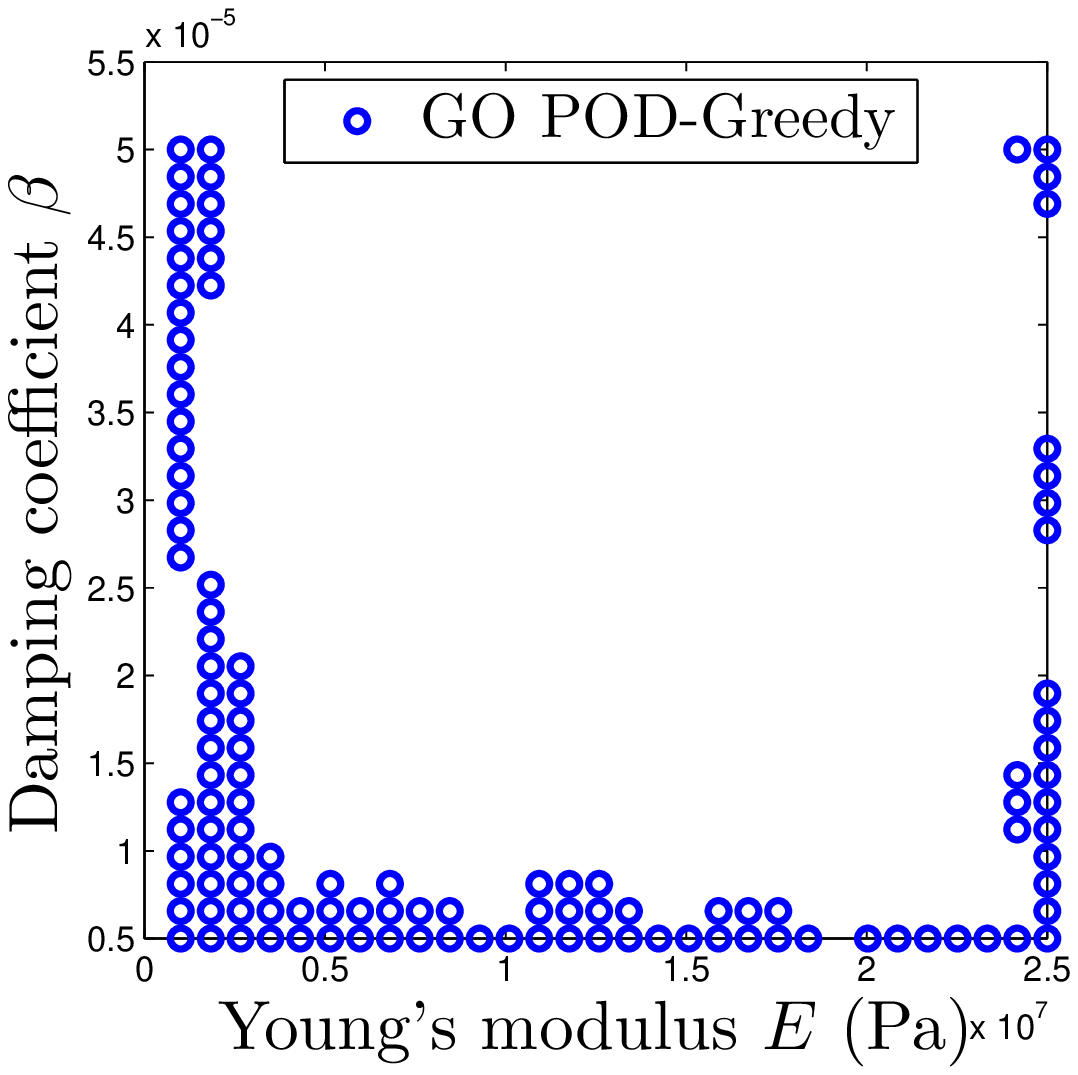}} \label{fig7b} }%
\caption{(a) Maximum of error indicator $\Delta_s(\mu)$ over $\Xi_{\rm train}$ as a function of $N$, and (b) distribution of sampling points by the goal-oriented POD--Greedy algorithm.}%
\label{fig7}
\end{center}
\end{figure}

\begin{figure}[h!]
\begin{center}
\subfigure[]{
\resizebox*{7cm}{!}{\includegraphics{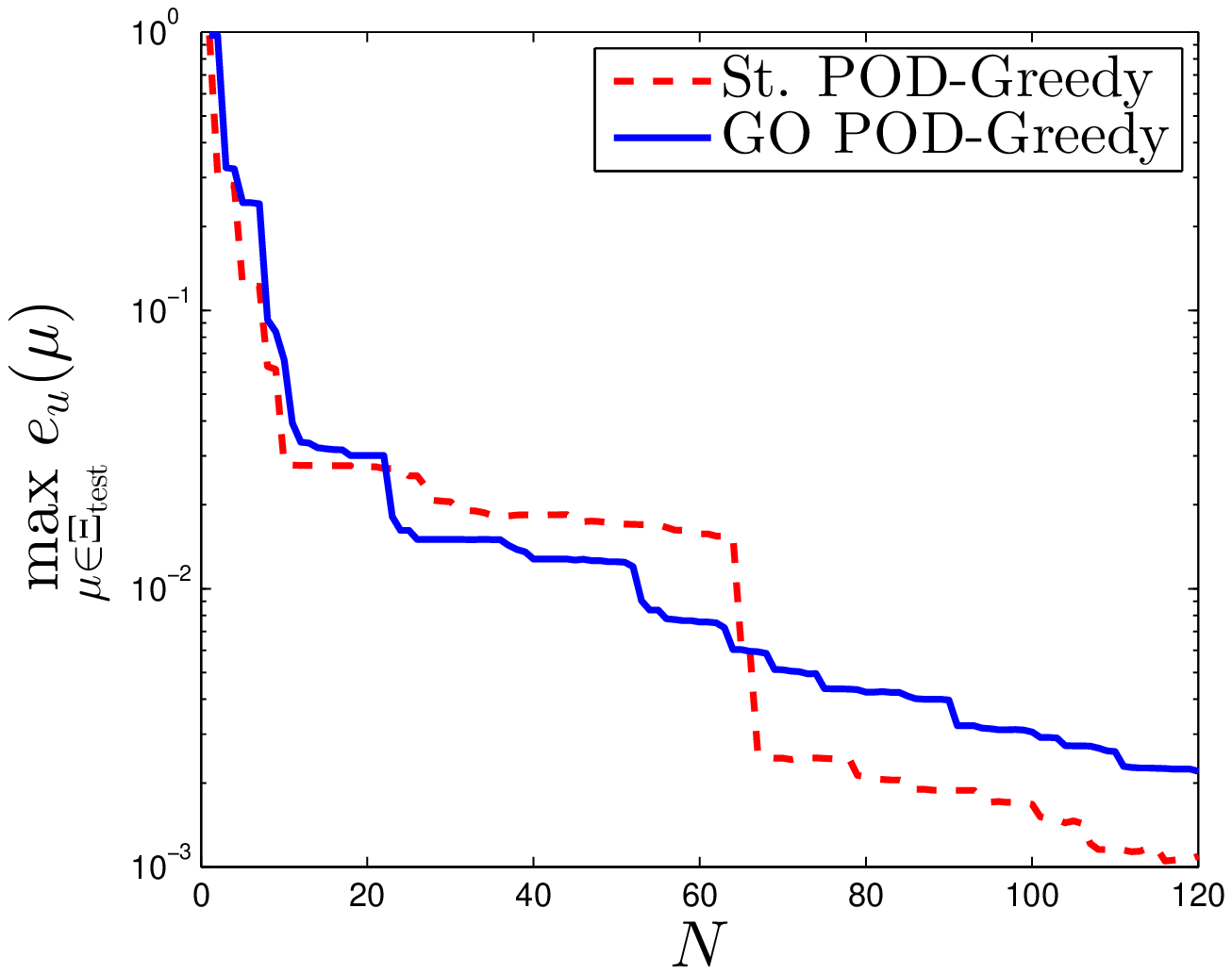}} \label{fig8a} }%
\subfigure[]{
\resizebox*{7cm}{!}{\includegraphics{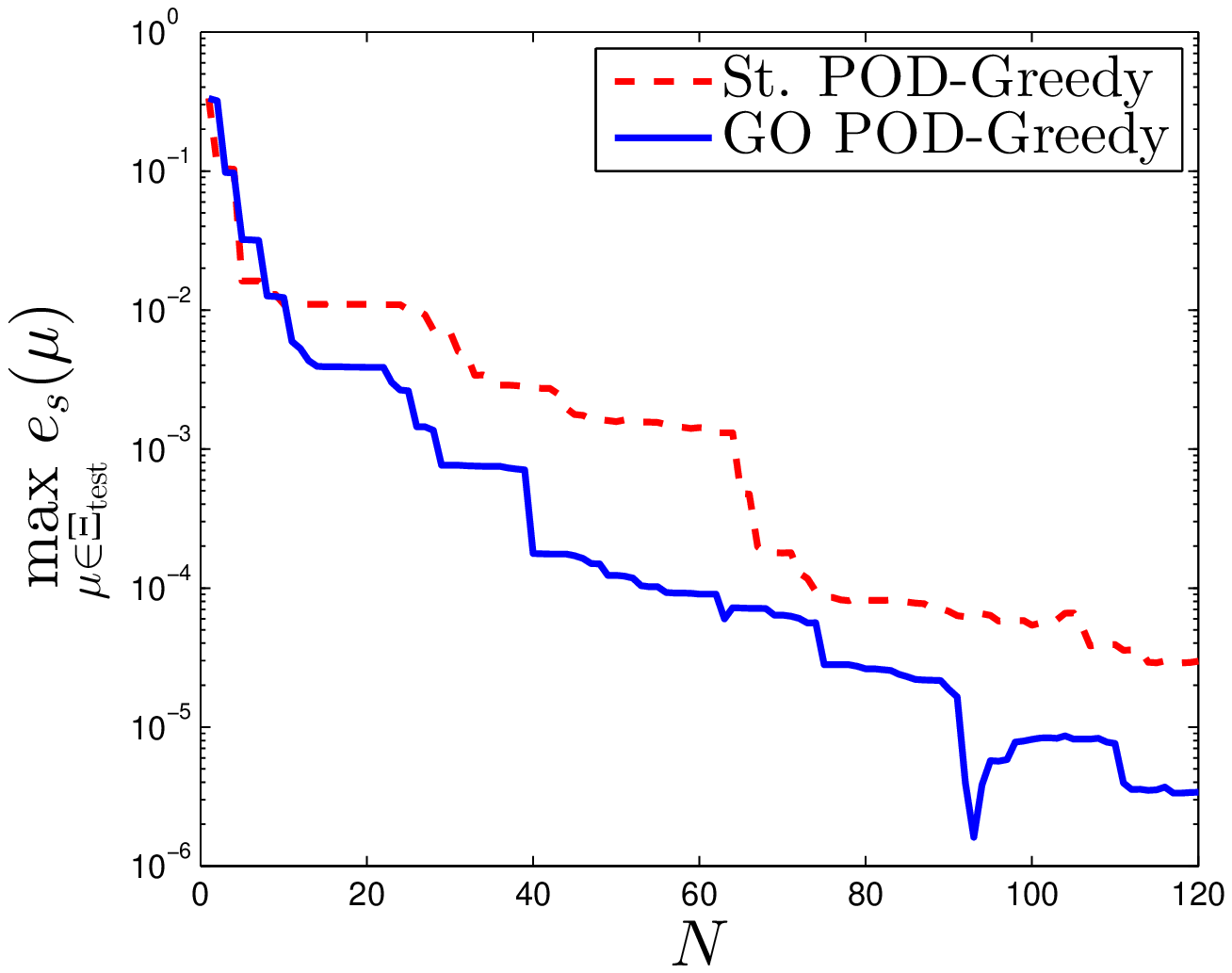}} \label{fig8b} }%
\caption{Comparison of maximum RB true errors by standard and goal-oriented POD--Greedy algorithms: (a) solution and (b) output over $\Xi_{\rm test}$.}%
\label{fig8}
\end{center}
\end{figure}

\begin{figure}[h!]
\begin{center}
\subfigure[]{
\resizebox*{7cm}{!}{\includegraphics{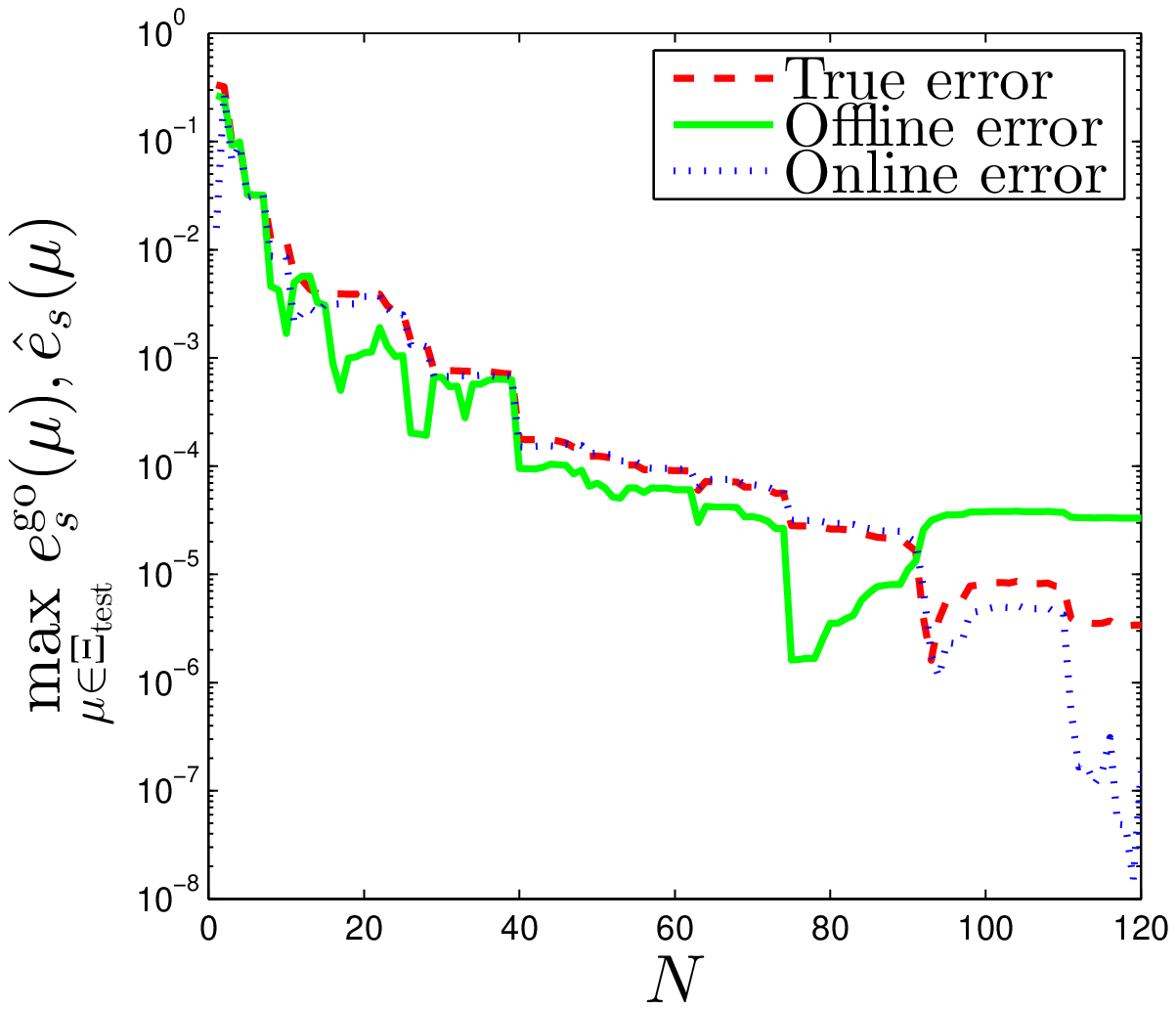}} \label{fig9a} }%
\subfigure[]{
\resizebox*{7cm}{!}{\includegraphics{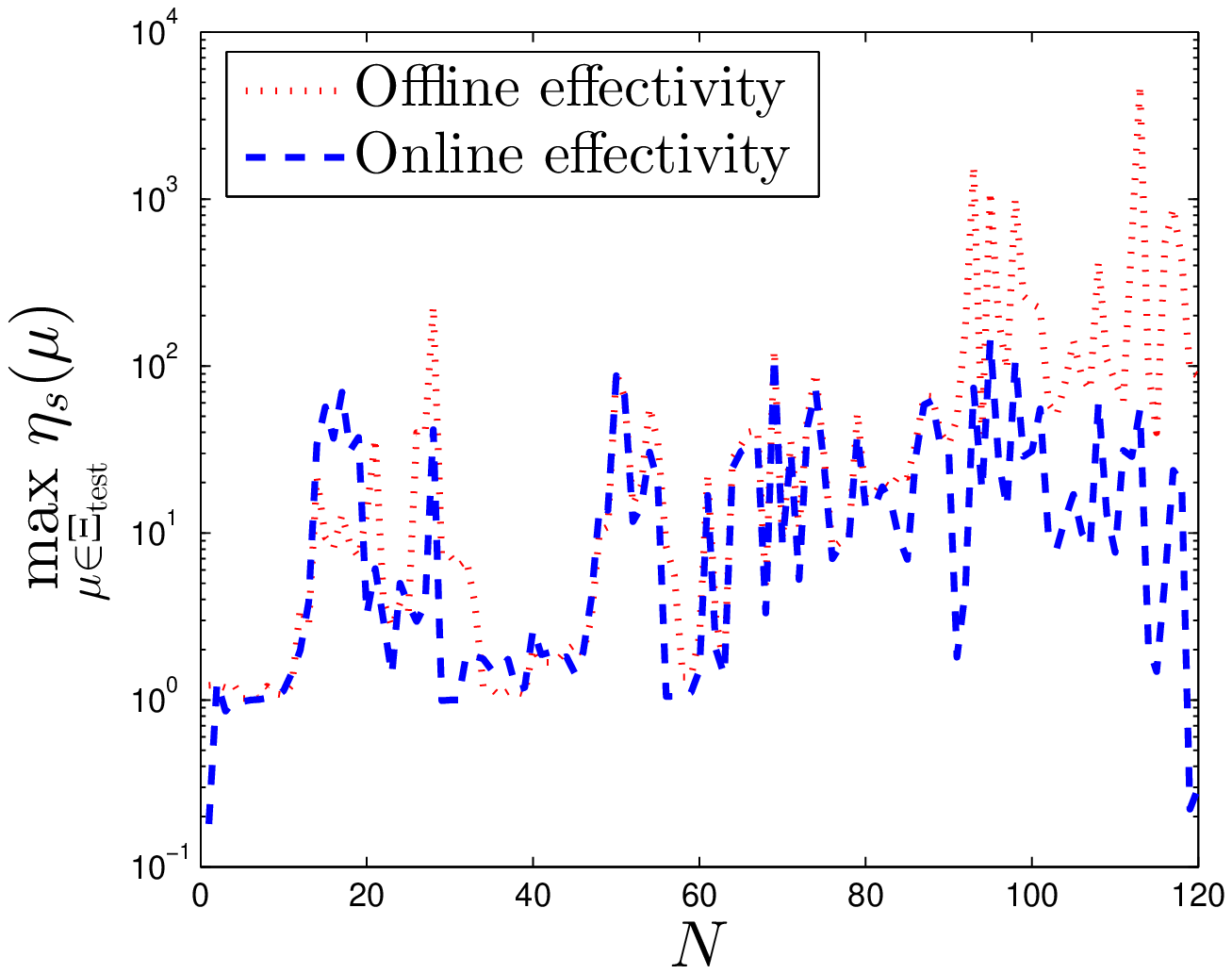}} \label{fig9b} }%
\caption{ (a) Comparison of max true output error versus max asymptotic output errors, and (b) their corresponding max effectivities over $\Xi_{\rm test}$. }%
\label{fig9}
\end{center}
\end{figure}

\begin{itemize}
  \item First, we implement the standard POD--Greedy algorithm (i.e., the left column of Table \ref{tab1}) \cite{Hoang2013}. The results are presented in Fig.\ref{fig6}: Fig.\ref{fig6a} shows the maximum error indicator $\Delta^{\max}_{u} = \max \limits_{\mu \in \Xi_{\rm train}} \Delta_u(\mu)$ as a function of $N$; and Fig.\ref{fig6b} shows the distribution of the sample set $S^{\rm st}_{*}$, respectively.

  \item Once the sets $S^{\rm st}_{*}$ and $Y^{\rm st}_{N}$ are available, the term $s^{\rm st}_{2N}(\mu)$ (of $\Delta_s(\mu)$ in (T\ref{tab1}j)) is now computable. Thus, we can now implement the goal-oriented POD--Greedy algorithm (i.e., the right column of Table \ref{tab1}). Consider the term $s^{\rm st}_{2N}(\mu)$, the subscript ``$2N$'' means that we approximate the FE output $s(\mu)$ by the (standard algorithm) enriched RB approximation $s^{\rm st}_{2N}(\mu)$ with $2N$ basis functions corresponding to the current $N^{\rm th}$-Greedy iteration of the goal-oriented algorithm: $N^{\rm st} = 2 N^{\rm go}$. Since $N^{\rm st}_{\max} = 120$, then we set

      \begin{equation}\label{eq:NumEx_two}
      N^{\rm st} =
          \begin{cases}
            2N^{\rm go} & \text{if } \quad 1 \le N^{\rm go} \le 60, \\
            120         & \text{if } \quad 60 < N^{\rm go} \le 120,
          \end{cases}
      \end{equation}

      to compute the term $s^{\rm st}_{2N}(\mu)$. Implementing the goal-oriented algorithm, the results are presented in Fig.\ref{fig7}: Fig.\ref{fig7a} shows the maximum error indicator $\Delta^{\max}_{s} = \max \limits_{\mu \in \Xi_{\rm train}} \Delta_s(\mu)$ as a function of $N$; and Fig.\ref{fig7b} shows the distribution of the sample set $S^{\rm go}_{*}$, respectively.

  \item In order to compare the performances of the two algorithms, we create the test sample set $\Xi_{\rm test} \subset \mathcal{D}$ which is a coarse subset of $\mathcal{D}$ and then test the two algorithms based on this set. The set is defined as follows: $\Xi_{\rm test} \equiv [1 \times 10^6, 25 \times 10^6]{\rm Pa} \times [5 \times 10^{-6}, 5 \times 10^{-5}]$ where $n_{\rm test}=100$ sample points distributed uniformly. We show, as functions of $N$: $e^{\max}_{u} = \max \limits_{\mu \in \Xi_{\rm test}} e_u(\mu)$ (defined in \eqref{eq:RB_five}) and $e^{\max}_{s} = \max \limits_{\mu \in \Xi_{\rm test}} e_s(\mu)$ (defined in \eqref{eq:RB_six}) in Fig.\ref{fig8}, respectively. From Fig.\ref{fig8a}, we see that the solution errors by the standard algorithm is a bit smaller than that of the goal-oriented one (with $N \ge 65$); while from Fig.\ref{fig8b}, the goal-oriented algorithm completely prevail over the the standard algorithm. The difference in the RB true output error by the two algorithms can be up to 10 times.

  \item Lastly, as regards error estimation, we show the plots of $\hat{e}^{\rm off,max}_{s} = \max \limits_{\mu \in \Xi_{\rm test}} \hat{e}^{\rm off}_{s} (\mu)$ (defined in \eqref{eq:RB_seven-a}), $\hat{e}^{\rm onl,\max}_{s} = \max \limits_{\mu \in \Xi_{\rm test}} \hat{e}^{\rm onl}_{s} (\mu)$ (defined in \eqref{eq:RB_seven-b}), and $e^{\rm go,\max}_{s} = \max \limits_{\mu \in \Xi_{\rm test}} e^{\rm go}_{s} (\mu)$ (defined in \eqref{eq:RB_six-b}) as functions of $N$ together in Fig.\ref{fig9a}. The maximum associated effectivities of these offline/online errors (i.e., $\eta^{\rm off,\max}_{s} = \max \limits_{\mu \in \Xi_{\rm test}} \eta^{\rm off}_{s}(\mu)$ and $\eta^{\rm onl,\max}_{s} = \max \limits_{\mu \in \Xi_{\rm test}} \eta^{\rm onl}_{s}(\mu)$) are also shown in Fig.\ref{fig9b}, respectively. As observed from Fig.\ref{fig9}, we see that the online asymptotic output error \eqref{eq:RB_seven-b} is very close to the true output error \eqref{eq:RB_six-b} and that its maximum effectivity is around the range $[10-100]$ which is acceptable for a 3D time-dependent dynamic problem.

\end{itemize}

Finally, regarding computational time, all computations were performed on a desktop Intel(R) Core(TM) i7-3930K CPU @3.20GHz 3.20GHz, RAM 32GB, 64-bit Operating System. The computation time for the RB solver\footnotemark \footnotetext{The work focuses on real-time context with many online computations, the offline stage is done once and expensive: the total computational time is approximately 2 weeks (including all FEM solutions/outputs and RB true errors of the standard and goal-oriented algorithms) on this computer.} ($t_{\rm RB(online)}$), the CPU-time for the FEM solver by our code ($t_{\rm FEM}$) and the CPU-time saving factor $\kappa=t_{\rm FEM}/t_{\rm RB(online)}$ are listed on Table \ref{tab3}, respectively. We see that the RB solver is approximately $O(1000)$ times faster than the FEM solver; and thus it is clear that the RB is very efficient and reliable for solving time-dependent dynamic problems.

\begin{table}[h!]
\center
\caption{Comparison of the CPU-time for a FEM and RB analysis.}
{\begin{tabular}{|c|c|c|c|}
\hline \rule{0pt}{2.5ex}
    N & $t_{\rm RB(online)}$ ({\rm sec}) & $t_{\rm FEM}$ ({\rm sec}) & $\kappa=t_{\rm FEM}/t_{\rm RB(online)}$ \\ [0.5ex]
\hline \rule{0pt}{2.5ex} \hspace{-2.5mm}
   10 & 0.0072 & 29  & 4027 \\
   20 & 0.0081 & 29  & 3580 \\
   30 & 0.0106 & 29  & 2736 \\
   40 & 0.0140 & 29  & 2070 \\
   50 & 0.0243 & 29  & 1193 \\
   60 & 0.0300 & 29  & 966 \\
\hline
\end{tabular}}
\label{tab3}
\end{table}

\section{Conclusion}\label{sec_Conclusion}

A new goal-oriented POD--Greedy sampling algorithm was proposed. The proposed algorithm cooperates and improves further the standard POD--Greedy algorithm by using the asymptotic output error rather than the dual norm of residual as error indicator in the Greedy iterations. It is demonstrated that this type of error indicator will guide the Greedy iterations to select the parameter samples/points to optimize the true output error. The proposed algorithm is then verified by investigating a 3D dental implant problem in the time domain. In comparison with the standard algorithm, we conclude that our proposed algorithm performs better -- in terms of output's accuracy, and quite similar -- in terms of solution's accuracy. The proposed algorithm is applicable to any (regular) output functional and is thus very suitable within the goal-oriented RB approximation context.


\section*{Acknowledgements}

We are sincerely grateful for the financial support of the European Research Council Starting Independent Research Grant (ERC Stg grant agreement No. 279578) entitled ``Towards real time multiscale simulation of cutting in non-linear materials with applications to surgical simulation and computer guided surgery''.

\bibliographystyle{unsrt}        

\bibliography{myrefs_s3}

\end{document}